\documentclass[12pt]{article}

\usepackage[margin=1in]{geometry}
\usepackage[nosort]{cite}
\usepackage{amsmath,amsthm,amsfonts,hyperref, amssymb}

\usepackage{array,multirow}

\usepackage[matrix,arrow,frame,import,curve,color]{xy}

\newtheorem{lemma}{Lemma}
\newtheorem{result}{Result}

\numberwithin{equation}{section}
\numberwithin{table}{section}

\def\ord{{{\text{ord}}}}

\def\C{{\mathbb C}}
\def\cC{{\cal C}}

\def\cO{{\cal O}}
\newcommand\Deltah{\widehat{\Delta}}
\newcommand\Deltat{\widetilde{\Delta}}

\newcommand\fh{\widehat{f}}
\newcommand\gh{\widehat{g}}
\newcommand\at{\widetilde{a}}
\newcommand\bt{\widetilde{b}}

\newcommand\dt{\widetilde{d}}

\newcommand\ft{\widetilde{f}}
\newcommand\gt{\widetilde{g}}
\newcommand\ut{\widetilde{u}}
\newcommand\vt{\widetilde{v}}
\newcommand\wt{\widetilde{w}}

\def\g{{\mathfrak{g}}}
\def\e{{\mathfrak{e}}}
\def\f{{\mathfrak{f}}}

\def\rep#1{{{\boldsymbol{#1}}}}

\def\ff#1#2{{\textstyle\frac{#1}{#2}}}
\def\half{\frac{1}{2}}

\def\SO{\operatorname{SO}}

\def\SU{\operatorname{SU}}
\def\GU{\operatorname{U{}}}
\def\Sp{\operatorname{Sp}}

\def\so{\operatorname{\mathfrak{so}}}

\def\sp{\operatorname{\mathfrak{sp}}}

\def\su{\operatorname{\mathfrak{su}}}

\def\P{{\mathbb P}}

\begin{document}

\title{On the global symmetries of 6D superconformal \\field
theories}

\author{Marco Bertolini$^{1,2,3}$\footnote{e-mail: {\tt mb266@phy.duke.edu}} , 
Peter R. Merkx$^4$\footnote{e-mail: {\tt merkx@math.ucsb.edu}} , 
and David R. Morrison$^{4,5}$\footnote{e-mail: {\tt drm@math.ucsb.edu}}
}
\date{}

\maketitle

\vspace*{-6.5cm}
\begin{flushright}
{NSF-KITP-15-141}
\end{flushright}

\vspace*{5.2cm}

{\small

{}$^1$ Center for Geometry and Theoretical Physics, Duke University,
Durham NC 27708 \vspace{2mm}

{}$^2$ Kavli Institute for Theoretical Physics, U.C. Santa Barbara, Santa Barbara CA 93106 \vspace{2mm}

{}$^3$ Perimeter Institute for Theoretical Physics, 31 Caroline Street North, ON N2L 2Y5, Canada \vspace{2mm}

{}$^4$ Department of Mathematics, U.C. Santa Barbara, Santa Barbara CA 93106 \vspace{2mm}

{}$^5$ Department of Physics, U.C. Santa Barbara, Santa Barbara CA 93106

}

\bigskip
\bigskip

\begin{abstract}
We study global symmetry groups of six-dimensional superconformal field theories (SCFTs).
In the Coulomb branch we use field theoretical arguments to predict an
upper bound for  the global symmetry of the SCFT.
We then analyze global symmetry groups of F-theory constructions of SCFTs with a one-dimensional Coulomb branch.
While in the vast majority of cases, all of the global symmetries allowed
by our Coulomb branch analysis can be realized in F-theory, in a handful
of cases we find that F-theory models fail to realize the full symmetry
of the theory on the Coulomb branch.  In one particularly mysterious
case, F-theory models realize several distinct maximal subgroups of the
predicted group, but not the predicted group itself.
\end{abstract}

\newpage

\tableofcontents

\section{Introduction}

The discovery of nontrivial six-dimensional superconformal field theories
(SCFTs)
nearly twenty years ago 
\cite{Witten:1995zh,SeibergWittensix,Seiberg:1996qx}
was quite surprising.
On the one hand,
it was known that such theories could exist in principle \cite{Nahm:1977tg},
but on the other hand it was argued that they could not be {Lagrangian}\/
theories, {\it i.e.}, that they could not 
 be studied as perturbations of free field theory.

Instead, a variety of indirect constructions were given, all
derived from string
theory and its cousins:  theories of point-like heterotic
instantons \cite{Witten:1995gx,GanorHanany,SeibergWittensix}
(sometimes located at singularities \cite{Aspinwall:1996vc,instK3}), 
theories on five-branes (sometimes located at singularities
\cite{Intriligator:1997kq,Blum:1997fw,Blum:1997mm}),
theories built from NS-branes and D-branes in 
type IIA \cite{Brunner:1997gk,Hanany:1997sa,Brunner:1997gf}, 
theories built from local F-theory 
models \cite{FCY1,FCY2,Bershadsky:1996nu,Aspinwall:1996vc,Bershadsky:1997sb,instK3}, 
and so on.
More recently, AdS$_7$ duals have been found for some infinite
families of these theories \cite{Apruzzi:2013yva,Gaiotto:2014lca}.  
Examples have been studied with either $(2,0)$ or $(1,0)$ supersymmetry;
in this paper, we focus on theories
with minimal $(1,0)$ supersymmetry in six dimensions.

It is generally difficult to directly
identify the degrees of freedom in these theories.  One important clue
is provided by the global symmetry group of the theory, together with
its action on the fields of the theory, so it is desirable
to determine this group whenever possible.\footnote{Knowing the
global symmetry group may also aid in identifying a possible $AdS_7$
dual for a family of theories.}  In many of the string-related
constructions of these theories, the global symmetry group is manifest,
but in one key case---constructions from F-theory---the global
symmetry group is difficult to determine.  In this paper, we investigate
global symmetry groups of these F-theory constructions, determining
them in some cases and constraining them in others.

The motivation for focusing on this particular type of six-dimensional
SCFTs comes from  recent dramatic progress in classifying
such theories \cite{6D-SCFT,atomic,6D-flows}.\footnote{As 
has been pointed out to us privately by several
different people, the classification of \cite{6D-SCFT,atomic,6D-flows}
implicitly assumes that there are no ``frozen'' seven-branes 
\cite{Witten:1997bs,triples,Tachikawa:2015wka}
in the F-theory models. Inclusion
of such seven-branes might possibly expand the list of known theories.}
The classification result of \cite{6D-SCFT} shows that for any
6D SCFT constructed from F-theory and having a Coulomb branch, the base
of the F-theory model is an orbifold of the form $\mathbb{C}^2/\Gamma$,
so we begin in section \ref{sec:orbifold}
 with a discussion of F-theory on orbifold
bases.
We then turn in section \ref{sec:local} to a more general consideration of
the local F-theory constructions of 6D SCFTs,
explaining that the constructed theory is not directly conformal
but rather flows to an interacting conformal theory under renormalization.
In fact, the F-theory construction itself contains parameters which
are irrelevant in the conformal limit, and for special values of those parameters
a global symmetry group may be manifest.  However, the global symmetry group
is not manifest for generic values, and so in most cases the global symmetries
of the theory must be
viewed as  emergent
symmetries in the IR limit (from the F-theory perspective).

In section \ref{sec:Coulomb}, we study these theories in their 
Coulomb branch, where they take the form of a conventional 6D
supersymmetric field theory
with hypermultiplets, vector multiplets, and tensor multiplets.
All global symmetries of the SCFT act as global symmetries of
the field theory on the Coulomb branch, so the global symmetry group
of the latter constrains the global symmetry group of the SCFT.
For theories whose Coloumb branch contains one tensor multiplet,
we present these field theoretic global symmetry groups
in section \ref{sec:predictions}
(using results of \cite{Danielsson:1997kt,Bershadsky:1997sb,anomalies}),
in the case that the theory on the Coulomb branch is a gauge theory.

Finally, in the remainder of the paper, we determine the possible
manifest global symmetries of F-theory models whose Coulomb branch
contains precisely one tensor multiplet.  
In many cases, we
find that all of the global symmetries of the field theory can 
be realized in some particular F-theory model.  In some cases,
however, we show that this is not possible,\footnote{As this paper
was nearing completion, a beautiful paper appeared \cite{Ohmori:2015fk}
which showed on field theory grounds that in one of the cases for
which we find a mismatch between the Coulomb branch prediction and
the F-theory possibilities, the SCFT itself does not have the same
global symmetry as on the Coulomb branch, but instead has the restricted
global symmetry found in F-theory.} and we determine instead
the (relatively) maximal global symmetry
groups which can be realized in F-theory.
For one particularly mysterious SCFT, 
there is more than one such relatively maximal
subgroup of the field theoretic global symmetry group.
We summarize these results in subsection~\ref{subsec:summary}.

In section \ref{sec:final} we state our conclusions and outline open problems.

\subsection*{Acknowledgments}
   It is a pleasure to thank I.~Bah, C.~Beem, C.~C{\'o}rdova, J.~Distler,
T.~Dumitrescu, J.~Halverson, K.~Intriligator,
P.~Koroteev, R.~Plesser,
T.~Rudelius and W.~Taylor
for useful discussions.
This work is supported in part by National Science Foundation grants
PHY-1125915,
PHY-1217109,
PHY-1521053,
and PHY-1307513.
MB and DRM thank the organizers of the workshop on Heterotic Strings and (0,2) QFTs at Texas A\&M  University in May, 2014 for hospitality while some of this work was being carried out.
DRM also thanks the Simons Center for Geometry and Physics for hospitality during the latter stages of this project.

\section{F-theory on orbifolds} \label{sec:orbifold}

Recently, there has been effort in classifying six-dimensional superconformal field theories (SCFTs) \cite{6D-SCFT,atomic} from F-theory. 
The data necessary to define an F-theory model are given by specifying a compact base space $B$, a line bundle over $B$ 
and sections $f$ and $g$ of the fourth and sixth powers of this line bundle.
These data determine  the Weierstrass model of a fibration $\pi: X \rightarrow B$, and 
the discriminant $4f^3+27g^2$ of the Weierstrass model whose zeros
(the ``discriminant locus'') describe where the fibration is 
singular.\footnote{We are assuming that there are no ``frozen'' seven-branes
\cite{Witten:1997bs,triples,Tachikawa:2015wka}.}
To obtain a conventional F-theory model ({\it i.e.}, one
not involving tensionless strings), the base $B$ must be nonsingular and 
the multiplicities of $f$ and $g$ cannot simultaneously exceed $4$ and $6$ 
at any point of $B$.  But it is also possible to use a Weierstrass model
to directly study theories having tensionless strings.

It is convenient to generalize this construction by considering a local model 
for $B$, in other words, to relax the condition that $B$ be compact.  In this
case, if we focus on a particular point $P$ of $B$ or if we consider
a collection of curves $\Sigma_i\subset B$ which are contractible to a point,
there will  be a scaling limit in which gravity decouples and a conformal
fixed point emerges.  

In the case of a point $P$ in $B$, if 
the point is a nonsingular one at which the multiplicity of $f$ is
at most $3$ or the multiplicity of  $g$ is at most $5$, then the 
corresponding conformal
fixed point will simply be a free theory.  However, if the theory had
tensionless strings before rescaling, then one expects an interacting SCFT.
Something similar happens if the point $P$ is itself a singular point of $B$.

The classification results of \cite{6D-SCFT,atomic,6D-flows} begin with
an F-theory model in which the singularities of $B$ and any points for
which $(f,g)$ have high multiplicity have been blown up.  That is, the
starting point is the so-called\footnote{This is sometimes referred
to as the {\em tensor branch}.} {\em Coulomb branch}\/ of the SCFT, in 
which nonzero expectation values have been given to the tensor multiplets
in the theory.  Concretely, this amounts to choosing a nonsingular base $B$
and a collection of compact curves $\{\Sigma_i\}$ to serve as (some of)
the components of the discriminant locus, 
such that the collection of curves can be 
contracted to a point, which might be a singular point.
The basic idea is that when blowing down the compact curves to obtain the CFT, there are, in general, non-compact curves carrying 
non-abelian gauge algebras, which in the singular limit give rise to the global symmetry of the SCFT.  This identification of global symmetries goes
back to the early discussions of an F-theory realization of the
small $E_8\times E_8$ instanton \cite{FCY1,WitMF,FCY2}, and
has been used in a number of recent papers \cite{6D-SCFT,DelZotto:2014hpa,atomic}.

The classification result in \cite{6D-SCFT} shows that for any local
F-theory model of a 6D SCFT having a Coulomb branch,
the base (before blowing up)
must have at worst an orbifold singularity $\mathbb{C}^2/\Gamma$
for some $\Gamma \subset \GU(2)$.  
Since we are interested in studying the Weierstrass models of such theories
directly, we will explain in this section
how to formulate F-theory on such bases, reviewing and extending the discussion
in \cite{6D-gravity}.

For any base $B$, the data of a supersymmetric F-theory model are specified
by Weierstrass coefficients $f$ and $g$ which are sections of
$\mathcal{O}(-4K_B)$ and $\mathcal{O}(-6K_B)$, respectively.
Moreover, the total space of the elliptic fibration is the compactification of
a threefold inside
\[ \mathcal{O}(-2K_B)\oplus \mathcal{O}(-3K_B)~,\]
defined by an equation in Weierstrass form
\begin{equation}
 -y^2 + x^3 + fx + g=0~,
\label{eq:W}
\end{equation}
where $(x,y)$ are local coordinates taking values in
$\mathcal{O}(-2K_B)\oplus\mathcal{O}(-3K_B)$, where $K_B$ is the canonical class of $B$.
The challenge is
interpreting the sheaves $\mathcal{O}(-mK_B)$ when the base $B$ is allowed
to have orbifold singularities.

When $B$ is nonsingular, each $\mathcal{O}(-mK_B)$ is a line bundle,
{\it i.e.}, a locally free sheaf of rank one.  In the presence of orbifold
singularities, however, these sheaves may only be ``reflexive'' sheaves
of rank one (see, for example, \cite{Reid:ypg}).  
In practice, this means that while for any point $Q\ne P$
there is a section of the sheaf which does not vanish at $Q$, it may
be necessary to use several different sections to ensure that at least
one of them is nonzero at each point in a small neighborhood of $P$.
Once appropriate bases have been determined, the Weierstrass equation
can be interpreted as an equation with coefficients in the coordinate
ring of $B$.

We illustrate this point with the example of 
$\mathbb{C}^2/\mathbb{Z}_3$, with the generator of the group acting via
\[ (z,w)\mapsto (e^{2\pi i/3}z,e^{2\pi i/3}w)~.\]
On the complement of $z=0$, local coordinates on the quotient are
provided by $s=z^3$ and $t=w/z$ (with $s\ne0$).  On the other hand,
on the complement of $w=0$, local coordinates are provided by
$u=z/w$ and $v=w^3$ (with $v\ne0$).  The coordinate change
map is given by
\begin{align}
s &= u^3v~, \qquad &t&=\frac1u~,
\end{align}
which implies that $ds\wedge dt = u du\wedge dv$.

To find generating sections of $\mathcal{O}(-mK)$, we consider an expression
of the form 
\begin{align} 
\frac{s^it^j}{(ds\wedge dt)^{\otimes m}}
= \frac{u^{3i-j-m}v^i}{(du\wedge dv)^{\otimes m}}~.
\end{align}
It is apparent that this gives a regular section whenever $j\ge0$ and
$3i-j-m\ge0$.

If we consider the case of $m=0$, we are simply computing the invariant
functions.  These are generated by
\begin{align} 
&1~, \quad &\alpha &:= s = u^3v~, \quad &\beta &:= st = u^2v~, \quad &\gamma &:= st^2 = uv~,
\quad &\delta &:= st^3=v~.
\end{align}
There are several relations among these generators:  
\begin{align}
\beta^2=\alpha\gamma~, \qquad\qquad
\gamma^2=\beta\delta~, \qquad \text{ and } \qquad  \alpha\delta=\beta\gamma~,
\label{eq:relations}
\end{align}
and it is known
that the coordinate ring is
\begin{align} 
\mathcal{R}:=\mathbb{C}[\alpha,\beta,\gamma,\delta]/(\beta^2-\alpha\gamma,
\gamma^2-\beta\delta, \alpha\delta-\beta\gamma)~.
\end{align}

Now, a general element $g$ of $\mathcal{O}(-6K_B)$ is simply given by
\begin{align}
\frac{g_0s^2}{(ds\wedge dt)^{\otimes6}} = \frac{g_0v^2}{(du\wedge dv)^{\otimes6}}~,
\end{align}
where $g_0\in\mathcal{R}$.  (This is because $\mathcal{O}(-6K_B)$
is locally free.)  The form of a general element $f$ of $\mathcal{O}(-4K_B)$ 
is more complicated; it is given by 
\begin{align} 
\frac{f_0s^2+f_1s^2t+f_2s^2t^2}{(ds\wedge dt)^{\otimes4}}
= \frac{f_0u^2v^2+f_1uv^2+f_2v^2}{(du\wedge dv)^{\otimes4}}~,
\end{align}
where $f_0$, $f_1$, and $f_2$ are all elements of $\mathcal{R}$.
Note, however, that the $f_j$ are not uniquely determined by $f$; an
alteration which replaces $(f_0,f_1,f_2)$ by
\begin{equation}
(f_0 + \beta \varphi_1 + \gamma \varphi_2 + \delta \varphi_3,
f_1 - \alpha \varphi_1 - \beta (\varphi_2+\psi_1) - \gamma (\varphi_3+\psi_2)
- \delta \psi_3,
f_2 + \alpha \psi_1 + \beta \psi_2 + \gamma \psi_3)~,
\end{equation}
where $\phi_i,\psi_i \in \mathcal{R}$,
does not change $f$.

Writing the Weierstrass equation involves variable elements 
$x_0, x_1, y_0 \in \mathcal{R}$ with
\begin{align} 
x = \frac{x_0s+x_1st}{(ds\wedge dt)^{\otimes2}}
= \frac{x_0uv+x_1v}{(du\wedge dv)^{\otimes2}} \qquad
\text{ and } \qquad
y = \frac{y_0s}{(ds\wedge dt)^{\otimes3}}
= \frac{y_0v}{(du\wedge dv)^{\otimes3}}~,
\end{align}
representing arbitrary sections of $\mathcal{O}(-2K_B)$ and 
$\mathcal{O}(-3K_B)$, respectively.  Note that the pair
$(x_0,x_1)$ is
not uniquely determined by $x$, but that alterations of the form
\begin{equation}
(x_0,x_1) \mapsto (x_0 + \beta \xi_1 + \gamma \xi_2 + \delta \xi_3,
x_1-\alpha\xi_1-\beta\xi_2-\gamma\xi_3)~,
\label{eq:alteration}
\end{equation}
where $\xi_i$ are elements of $\mathcal{R}$,
leave $x$ unchanged.

Substituting into the usual Weierstrass equation \eqref{eq:W} gives
a section of $\mathcal{O}(-6K_B)$ which must vanish.  Writing this
as a multiple of the generator
\begin{align}
\frac{s^2}{(ds\wedge dt)^{\otimes6}} = \frac{v^2}{(du\wedge dv)^{\otimes6}}~
\end{align}
gives an equation in $\mathcal{R}$, namely,
\begin{equation}
y_0^2 = \alpha(x_0^3+f_0x_0) + \beta(3x_0^2x_1+f_1x_0+f_0x_1)
+\gamma(3x_0x_1^2 + f_2x_0+f_1x_1) + \delta(x_1^3+f_2x_1)+g_0 ~,
\end{equation}
which must be satisfied in addition to the 
relations \eqref{eq:relations} in order to determine
an elliptic fibration over $B$.  (Note that the equivalence relation
\eqref{eq:alteration} must also be imposed.)

\section{6D SCFTs from local F-theory models} \label{sec:local}

Most studies of 6D SCFTs constructed from F-theory have studied the models
on the Coulomb branch, where there are no tensionless strings and where
one thus expects a conventional field theory once gravity has been decoupled.

We propose instead that 6D SCFTs are most naturally studied on a singular
local model which has tensionless strings associated with a particular point
in the geometry, because this brings us directly into contact with
the conformal field theory.  The simplest example
 of this is the F-theory realization
of the small $E_8$ instanton, which consists of a point on the base $B$
at which the multiplicity of $f$ is $4$ and the multiplicity of $g$ is
$6$. Traditionally, this theory
is studied by blowing up the point to obtain the Coulomb branch of the
theory, and then rescaling the metric to zoom in on a neighborhood of
the curve (which is contracted back to a point in the process, if the
scaling is appropriately chosen).  Here, we suggest that this scaling
should be done directly on the singular F-theory model, zooming in on the
point $P$ supporting the singularity and only considering an appropriately
rescaled metric in a neighborhood of that point.
(This extends the analysis of \cite{Cordova:2009fg} to the case of
a noncompact base.)

The key features of F-theory at the point $P$ after
zooming are captured by the leading terms in the
Weierstrass equation, truncating
$f$ and $g$ appropriately:
\[ y^2 = x^3 + \tilde f(z,w)x + \tilde g(z,w)~,\]
where $\tilde f$ and $\tilde g$ contain only the low-order terms
of $f$ and $g$.  (Precisely which low-order terms are needed depends
on the type of superconformal fixed point.)
What is perhaps surprising is that varying the coefficients
 of the polynomials
$\tilde f$ and $\tilde g$, or including higher-order terms,
 does not affect the SCFT.  The coefficients must be sufficiently
generic to avoid converging to a different SCFT, and in particular cannot
be zero, but any generic choice will do.

One may wonder if this phenomenon has some five-dimensional interpretation
upon circle compactification.  It does not:  the new degrees of freedom
in the five-dimensional theory live in vector multiplets (and geometrically
correspond to the possibility of blowing up singularities in the total
space of the elliptic fibration).  However, this phenomenon of extra
polynomial coefficients not needed to describe the CFT actually 
corresponds to hypermultiplets in a compact model of F-theory which
would still be hypermultiplets upon circle compactification.  These
hypermultiplets decouple in the CFT limit.

Our interpretation of this observation is that the local F-theory data
are specifying more than just the SCFT, they are specifying 
an irrelevant deformation of the SCFT, with the coefficients in $f$ and $g$
choosing which irrelevant deformation is being made.\footnote{It is known
that 6D SCFTs do not admit marginal or relevant deformations, but can
admit irrelevant ones \cite{CDI}.}
In other words, the local F-theory data specify a point in the configuration
space for the field theories which lies to the ultraviolet of the SCFT
and flows to it under renormalization.

Some of the local F-theory realizations of a given SCFT may make the global
symmetries of the SCFT manifest, but this is not a requirement.  
When present, the
global symmetries are visible in an F-theory realization in the form
of a component of the discriminant locus $\{4f^3+27g^2=0\}$
which passes through the point $P$.  After rescaling, this component
is necessarily non-compact, and the coupling of the associated gauge
group has gone to zero, leaving a global symmetry.

For example, the particular realization of the small $E_8$ instanton
with equation
\begin{align}
y^2 = x^3 + z^4x + z^5w~,
\end{align}
has discriminant locus $\{z^{10}(4z^2+27w^2)=0\}$ and  has a manifest
$E_8$ global symmetry along $z=0$.  However, the generic realization
of this same SCFT
has {\em no}\/ manifest global symmetry.

In this paper, we shall determine the maximal manifest global symmetry
which can be obtained in F-theory realizations of SCFTs with a one-dimensional
Coulomb branch.  We do our calculations on the Coulomb branch, but our
conclusions are properly drawn after the distinguished
curve (created by blowing up)
is contracted.

\section{Global symmetries in the Coulomb branch} \label{sec:Coulomb}

All global symmetries of the SCFT act as global symmetries of
the field theory on the Coulomb branch, so the global symmetry group
of the latter constrains the global symmetry group of the 
SCFT.\footnote{It appears possible for a direct factor of the
global symmetry group of the SCFT to act trivially on the Coulomb branch,
but the only known cases where this might possibly happen are not
gauge theories, so we do not consider this possibility here.  
We will return to this point at the end of 
subsection~\ref{subsec:summary}.}
We determine these field theoretic global symmetry groups
(using results of \cite{Danielsson:1997kt,Bershadsky:1997sb,anomalies})
for gauge theories with a one-dimensional Coulomb branch.

As described in section \ref{sec:orbifold}, we specify an F-theory model 
in six dimensions via
sections $f$ and $g$ of appropriate rank-one reflexive sheaves on a base
$B$ which is a complex surface.  These sections in turn determine
an elliptically fibered Calabi-Yau threefold $\pi:X\rightarrow B$ with a 
section, typically singular, by means of a Weierstrass equation
\begin{align}
\label{eq:weierstr}
y^2 = x^3 + f x + g~.
\end{align}
This equation has an associated  discriminant 
\begin{align}
\Delta:=4f^3+27g^2~,
\end{align}
and discriminant locus
\begin{align}
\{\Delta =0\}~,
\end{align}
which indicates where the fibration is singular.\footnote{In the language of type IIB string theory, the discriminant locus indicates the location of seven-branes wrapping divisors on $B$.} 
These quantities are sections of certain line bundles
\begin{align}
\label{eq:fgDelta}
f& \leftrightarrow \cO(-4K_B)~,	&g&\leftrightarrow \cO(-6K_B)~, 	&\Delta&\leftrightarrow \cO(-12 K_B)~.
\end{align}
The above fibration corresponds to a non-singular F-theory model when it is possible to blow-up the singularities in \eqref{eq:weierstr} to obtain a smooth
elliptically fibered Calabi-Yau $\widetilde X$.  F-theory compactified on $\widetilde X$ will yield a six-dimensional theory, whose gauge group $G$ is determined by the singularity structure of the elliptic fibration. 
We will focus our analysis on the non-abelian part of the Lie algebra $\mathfrak{g}$, which is characterized by the codimension one singularities, and on the matter content of the theory, which is encoded in the codimension two singularities.
The degrees of vanishing of the quantities in \eqref{eq:fgDelta} along a curve $\Sigma$ suffice in most cases to determine the gauge algebra along $\Sigma$
according to the Kodaira classification \cite{Kodaira,Neron}.  In some cases, however, information regarding monodromy is needed to determine the precise
form of the gauge algebra.  This is summarized in table \ref{t:singtypes}.
\begin{table}[t]
\begin{center}
\begin{tabular}{ccc|c|c|c}
ord($f$)	&ord($g$)		&ord($\Delta$)		&type			&singularity			&non-abelian algebra \\
\hline
$\geq0$	&$\geq0$		&0				&I${}_0$			&none				& none \\
$0$		&$0$			&1				&I${}_1$			&none				& none \\
$0$		&$0$			&$n\geq2$		&I${}_n$			&$A_{n-1}$			&$\su(n)$ or $\sp([n/2])$ \\
$\geq1$	&$1$			&2				&II				&none				& none \\
$1$		&$\geq2$		&3				&III				&$A_1$				&$\su(2)$\\
$\geq2$	&$2$			&4				&IV				&$A_2$				&$\su(3)$ or $\su(2)$ \\
$\geq2$	&$\geq3$		&6				&I${}_0^\ast$		&$D_4$				&$\so(8)$ or $\so(7)$ or $\mathfrak{g}_2$ \\
$2$		&$3$			&$n\geq7$		&I${}_{n-6}^\ast$	&$D_{n-2}$			&$\so(2n-4)$ or $\so(2n-5)$\\
$\geq3$	&$4$			&8				&IV${}^\ast$		&$\mathfrak{e}_6$		&$\mathfrak{e}_6$ or $\mathfrak{f}_4$ \\
$3$		&$\geq5$		&9				&III${}^\ast$		&$\mathfrak{e}_7$		&$\mathfrak{e}_7$ \\
$\geq4$	&$5$			&10				&II${}^\ast$		&$\mathfrak{e}_8$		&$\mathfrak{e}_8$ \\
$\geq4$	&$\geq6$		&$\geq12$		&non-minimal		&-					&- 
\end{tabular}
\end{center}
\caption{Singularity types with associated non-abelian algebras.}
\label{t:singtypes}
\end{table}
The last entry in the table is designated ``non-minimal'' since after the singularities in the fibers are resolved, the resolution contains a curve which can be blown down, {\it i.e.}, it is not a ``minimal model'' in the sense of birational geometry.
In fact, blowing that curve down is associated to a new Weierstrass model in which the degrees
of vanishing of $(f,g,\Delta)$ have been reduced by $(4,6,12)$.
For this reason we need not consider such cases.
 
The ``monodromy'' which determines which gauge algebra occurs for a given
singularity type is part of Tate's algorithm \cite{MR0393039}, first discussed in 
the physics literature in \cite{geom-gauge}.  A general specification for
determining the gauge algebra was given in table 4 of \cite{anomalies}, which
we review here.  In each case, there is a covering of the curve
$\Sigma$ which can be described by means of an algebraic equation
in an auxiliary variable $\psi$ (which takes its values in a line
bundle over $\Sigma$).  If $\Sigma$ is defined by $z=0$, then these
monodromy covers are given as follows \cite{anomalies}:

\begin{center}
\begin{tabular}{|c|c|} \hline
type 
&equation of monodromy cover\\ \hline\hline
I${}_m$, $m\ge3$
&$\psi^2+(9g/2f)|_{z=0}$
\\ \hline 
IV
&$\psi^2-(g/z^2)|_{z=0}$
\\ \hline
I${}_0^*$
& $\psi^3+(f/z^2)|_{z=0}\cdot\psi+(g/z^3)|_{z=0}$
\\ \hline
I${}_{2n-5}^*$, $n\ge3$
&$\psi^2+\frac14(\Delta/z^{2n+1})(2zf/9g)^3|_{z=0}$
\\ \hline 
I${}_{2n-4}^*$, $n\ge3$
&$\psi^2+(\Delta/z^{2n+2})(2zf/9g)^2|_{z=0}$
\\ \hline 
IV${}^*$
& $\psi^2-(g/z^4)|_{z=0}$
\\ \hline 
\end{tabular}
\end{center}

For all cases except I${}_0^*$, the equation of the
monodromy cover takes the form
``$\psi^2 - \text{something}$'',
and this cover splits (leading to no monodromy) if and only if 
the expression ``something'' is a perfect square.  In the remaining
case I${}_0^*$, the monodromy cover equation defines a degree $3$ cover of
$\Sigma$, and one must analyze this further to determine if that cover is
irreducible ($\mathfrak{g}(\Sigma) = \mathfrak{g}_2$),
splits into two components ($\mathfrak{g}(\Sigma) = \mathfrak{so}(7)$),
or splits into three components ($\mathfrak{g}(\Sigma) = \mathfrak{so}(8)$).
Each of these possibilities is reflected in the behavior of the residual
discriminant, as we will now describe.

First, suppose that the monodromy cover splits completely
(the case of gauge algebra $\so(8)$). That is, suppose that
\begin{align}
\psi^3+(f/z^2)|_{z=0}\cdot\psi+(g/z^3)|_{z=0} =
(\psi - \alpha)(\psi - \beta)(\psi-\gamma)~,
\end{align}
with $\alpha+\beta+\gamma=0$.  It is easy to calculate the
residual discriminant $(\Delta/z^6)|_{z=0}$ in this case; it is
\begin{align}
 4(\alpha\beta+\beta\gamma+\gamma\delta)^3+27\alpha^2\beta^2\gamma^2
&= - (\alpha-\beta)^2(\beta-\gamma)^2(\gamma-\alpha)^2 \nonumber\\
&= -(\alpha-\beta)^2(\alpha+2\beta)^2(2\alpha+\beta)^2~.
\end{align}
This form of the residual discriminant is a necessary consequence
of having gauge algebra $\so(8)$.  Notice that the analysis of
\cite{anomalies} associates to the vanishing of the factors $\alpha-\beta$,
$\beta-\gamma$, and $\gamma-\alpha$ the three different $8$-dimensional
representations of $\so(8)$.

Second, suppose that the monodromy cover splits into a linear factor 
and a quadratic factor (the case of gauge algebra $\so(7)$).  That is,
suppose that
\begin{equation}
\psi^3+(f/z^2)|_{z=0}\cdot\psi+(g/z^3)|_{z=0}
= (\psi - \lambda) (\psi^2 + \lambda \psi + \mu)~.
\label{eq:partially-split}
\end{equation}
We can calculate the residual discrimimant $(\Delta/z^6)|_{z=0}$ as
\begin{align}
 4(\mu-\lambda^2)^3+27\lambda^2\mu^2 = (\mu+2\lambda^2)^2(4\mu-\lambda^2)~.
\end{align}
The factor $\varphi:=4\mu-\lambda^2$  is the discriminant
of $\psi^2 + \lambda \psi + \mu=0$, so it captures the ramification points
of the double cover.  (Moreover, if $\varphi$ is a square, then the
double cover splits and we are back to the case of $\so(8)$.)
The analysis of \cite{anomalies} then shows that
these intersection points are associated to the $7$-dimensional representation
of $\so(7)$, while the other points (the zeros of $\mu+2\lambda^2$) are
associated to the spinor representation.

Notice that if the monodromy cover splits completely in 
\eqref{eq:partially-split}, we may choose one
of the roots (say $\alpha$) to use as $\lambda$, and then we have
$\mu=\beta\gamma$.  It follows that 
\begin{align}
\mu+2\lambda^2 &= \beta\gamma+2\alpha^2=(2\alpha+\beta)(\alpha-\beta)~, \nonumber\\
4\mu-\lambda^2 &= 4\beta\gamma-\alpha^2 = -(\alpha+2\beta)^2
\end{align}
(using $\gamma=-\alpha-\beta$), which allows for easy comparison between
the two forms of the discriminant.

\section{Prediction from field theory}
\label{sec:predictions}

In this section we want to describe the global symmetries that we would expect from field theory on the Coulomb branch for gauge theories with a
one-dimensional Coulomb branch.
Let us consider a curve $\Sigma$ with gauge algebra $\g$.  The possible
gauge algebras and matter representations were determined from anomaly
cancellation in \cite{Danielsson:1997kt,Bershadsky:1997sb};
a convenient place to find these results in F-theory language is
table 10 of \cite{anomalies}.\footnote{Our notation for representations 
follows \cite{anomalies} and is fairly standard:
for $\mathfrak{su}(n)$ and $\mathfrak{sp}(n)$,  $F$ denotes the
fundamental representation and $\Lambda^k$ denotes its exterior powers.
Note that for $\mathfrak{g}=sp(n)$, the representation $\Lambda^2$
contains a one-dimensional summand; the remainder is denoted by
$\Lambda^2_{irr}$.  For $\mathfrak{so}(n)$, $V$ denotes the vector 
representation, $S_+$ and $S_-$ denote half-spin representations,
and $S_\ast$ (according to \cite{anomalies}) denotes a spin or half-spin representation, depending on
the parity of $n$.  For the exceptional algebras, representations
are denoted by their dimension (in bold face type).}
Field theory predicts (see, for example, \cite{Gaiotto:2009we}) 
that $N$ hypermultiplets in a complex representation of
the gauge group support an $\SU(N)$ symmetry group, 
which is enhanced to $\SO(2N)$ for quaternionic ({\it i.e.}, pseudoreal) representations, and to $\Sp(N)$ 
for real representations.\footnote{Our convention is that $\Sp(N)$ is a 
subgroup of $\SU(2N)$.}  In the quaternionic case, the underlying complex
representation is a representation by half-hypermultiplets, so that $N$
is allowed to be a half-integer.  (We will emphasize this by declaring that
the representation consists of $M$ copies of $\frac12 V$, where
$M$ is allowed to be an arbitrary integer, leading to global symmetry
$\SO(M)$.)

Our conventions here differ slightly with the ones used in \cite{anomalies}:
to compare the two,
one needs to substitute $-K_B$ for $L$ in \cite{anomalies}.
Moreover, we interpret $(\dots)\big|_\Sigma$
as an intersection number $(\dots)\cdot \Sigma$, and do not worry about the 
algebraic cycle class discussed in \cite{anomalies}.  
Also, we note that the adjoint representation
never occurs in our models
since $\Sigma$ is rational, and thus we will simply substitute
\begin{align}
L\big|_\Sigma = - K_B\cdot \Sigma = 2 + \Sigma^2~,
\label{eq:KSigma}
\end{align}
into the formulas of \cite{anomalies}.
Note that \cite{anomalies} would have included 
$(1+\Sigma^2)\Lambda^2_{\text{irr}}$ as part of the representation content
of $\sp(n),\ n\geq2$, but since $\Sigma^2<0$ we conclude that $\Sigma^2=-1$
and this representation does not occur.

\begin{table}[t!]
\begin{tabular}{c|c|c}
$\g$					&representation										&global symmetry\\
\hline
$\su(2)$				&$(32+12\Sigma^2)\half F$								&$\so(32+12\Sigma^2)$\\
$\su(3)$				&$(18+6\Sigma^2)F$									&$\su(18+6\Sigma^2)$\\
$\su(4)$				&$(16+4\Sigma^2)F+(2+\Sigma^2)\Lambda^2$					&$\su(16+4\Sigma^2)\oplus\sp(2+\Sigma^2)$\\
$\su(5)$				&$(16+3\Sigma^2)F+(2+\Sigma^2)\Lambda^2$					&$\su(16+3\Sigma^2)\oplus\su(2+\Sigma^2)$\\
$\su(6)$				&$(16+2\Sigma^2)F+(2+\Sigma^2)\Lambda^2$					&$\su(16+2\Sigma^2)\oplus\su(2+\Sigma^2)$\\
$\su(6)^\ast$			&$(16+\Sigma^2)F+\half(2+\Sigma^2)\Lambda^3$				&$\su(16+\Sigma^2)\oplus\so(2+\Sigma^2)$\\
$\su(n), \ n\geq7$		&$(16+(8-n)\Sigma^2)F+(2+\Sigma^2)\Lambda^2$				&$\su(16+(8-n)\Sigma^2)\oplus\su(2+\Sigma^2)$\\
$\sp(n), \ n\geq2$		&$(16+4n)\half F$ 		&$\so(16+4n)$\\
$\so(7)$				&$(3+\Sigma^2)V+2(4+\Sigma^2)S_\ast$						&$\sp(3+\Sigma^2)\oplus\sp(8+2\Sigma^2)$\\
$\so(8)$				&$(4+\Sigma^2)V+(4+\Sigma^2)(S_++S_-)$					&$\sp(4+\Sigma^2)\oplus\sp(4+\Sigma^2)\oplus\sp(4+\Sigma^2)$\\
$\so(9)$				&$(5+\Sigma^2)V+(4+\Sigma^2)S_\ast$						&$\sp(5+\Sigma^2)\oplus\sp(4+\Sigma^2)$\\
$\so(10)$				&$(6+\Sigma^2)V+(4+\Sigma^2)S_\ast$						&$\sp(6+\Sigma^2)\oplus\su(4+\Sigma^2)$\\
$\so(11)$				&$(7+\Sigma^2)V+(4+\Sigma^2)\half S_\ast$					&$\sp(7+\Sigma^2)\oplus\so(4+\Sigma^2)$\\
$\so(12)$				&$(8+\Sigma^2)V+(4+\Sigma^2)\half S_\ast$					&$\sp(8+\Sigma^2)\oplus\so(4+\Sigma^2)$\\
$\so(13)$				&$(9+\Sigma^2)V+(2+\half\Sigma^2)\half S_\ast$					&$\sp(9+\Sigma^2)\oplus\so(2+\half\Sigma^2)$\\
$\so(n),\ n\geq14$		&$(n-8)V$												&$\sp(n-8)$\\
$\e_6$				&$(6+\Sigma^2)\rep{27}$									&$\su(6+\Sigma^2)$\\
$\e_7$				&$(8+\Sigma^2)\half \rep{56}$								&$\so(8+\Sigma^2)$\\
$\e_8$				&none												&none\\
$\f_4$				&$(5+\Sigma^2)\rep{26}$									&$\sp(5+\Sigma^2)$\\
$\g_2$				&$(10+3\Sigma^2)\rep{7}$								&$\sp(10+3\Sigma^2)$
\end{tabular}

\bigskip

\centerline{Note: $\Sigma^2=-1$ for  $\su(6)^\ast$;
$\Sigma^2=-1$ for  $\sp(n),\ n\geq2$;
$\Sigma^2=-4$ for $\so(n),\ n\geq14$; and
$\Sigma^2=-12$ for $\e_8$.}
\caption{Global symmetries as predicted from field theory.}
\label{t:fieldthypreds}
\end{table}

One of the observations of \cite{anomalies} is that in some cases there can be more than one matter representation with the same effect on anomaly
cancellation.  In the case of $\su(6)$ and $\Sigma^2=-1$,
$\Lambda^3$ plus two fundamentals (denoted by $\su(6)^\ast$) has the same anomaly content as
two copies of $\Lambda^2$ (denoted by $\su(6)$). As discussed in \cite{anomalies,matter1}, these different matter representations arise in enhancements from $A_5$ to $E_6$.
However, it is easy to see that for any other gauge group and a single contractible rational curve, none of these ``alternate" possibilities can occur.  In fact, table 3 of \cite{anomalies}
shows that in order to get one of these alternate possibilities, there must be at least two copies of $\Lambda^2$ or $\Lambda^2_{\text{irr}}$ (for $\su(n)$
or $\sp(n)$, respectively).   As we can read off from table~\ref{t:fieldthypreds}, this representation does not appear for $\sp(n)$, while for $\su(n)$
there are $2+\Sigma^2$ copies of $\Lambda^2$, but since $\Sigma^2<0$ this
means that there is at most one copy. This argument does not apply to $\su(6)$ because in this case $\Lambda^3$ is a pseudo-real representation.

A key point we want to emphasize is that the global symmetry only depends on the gauge algebra supported on $\Sigma$, and not on the Kodaira type
that realizes it. For example, we can realize $\su(2)$ with Kodaira types I${}_2$, I${}_3$, III and IV (with appropriate monodromies), and in all cases
the global symmetry is $\so(32+12\Sigma^2)$.  For this reason table \ref{t:fieldthypreds} lists only the algebras and not the Kodaira types. 

We have a minor correction, or reinterpretation, to offer concerning
table 10 of \cite{anomalies}: for $\so(8)$, $S_\ast$ must be
interpreted as $\frac12(S_++S_-)$.\footnote{In fact, this change
could be made uniformly for all $\so(2n)$ cases if desired, since
$S_+$ is Casimir equivalent to $S_-$ in all cases except $2n=8$.}
Also, the analysis in example 13 of \cite{anomalies} shows that $4+\Sigma^2$ must be divisible by 2 for $\so(13)$ and must be divisible by 4 for $\so(14)$.
Thus, $\Sigma^2=-4$ for $\so(14)$ and $\Sigma^2=-2,-4$ for $\so(13)$.
In addition, as explained in \cite{anomalies}, for Kodaira type
I${}_m^*$, $m\ge4$ (which corresponds
to $\so(n),\ n\ge15$), we must have $(-2K_B-\Sigma)\big|_\Sigma=0$, {\it i.e.},
$\Sigma^2=-4$.  Similarly, for Kodaira type II${}^*$ (which corresponds
to $\e_8$), $(-6K_B-5\Sigma)\big|_\Sigma=0$, {\it i.e.}, $\Sigma^2=-12$.

The reality conditions for $\su(n)$ are slightly subtle: the fundamental representation is quaternionic for $n=2$ and is complex for $n>2$; 
the anti-symmetric representation is real for $n=4$ and is complex for $n>4$.  In the $\sp(n)$ case the situation is simpler, as the fundamental 
representation is always quaternionic and the irreducible part of the anti-symmetric representation (which does not occur) is always real.

We also used the following reality conditions for the spinor representation(s) of $\so(n)$:
\begin{itemize}
\item If $n=0 \mod 8$, the two spinor representations are real.
\item If $n=1,7 \mod 8$, the unique spinor representation is real.
\item If $n=2,6 \mod 8$, the two spinor representations are complex.
\item If $n=3,5 \mod 8$, the unique spinor representation is quaternionic.
\item If $n=4 \mod 8$, the two spinor representations are quaternionic.
\end{itemize}

It is amusing to verify that table \ref{t:fieldthypreds} respects the exceptional isomorphisms of Lie algebras.  If we extend the $\so(n)$ formula to the 
representation of $\so(6)$, we find $(2+\Sigma^2)V + 4(4+\Sigma^2)S_\ast$, which agrees with the entry for $\su(4)$ once we identify 
$(V,S_\ast)$ with $(\Lambda,F)$.  Similarly, in the case of $\so(5)$, we find $(1+\Sigma^2)V+8(4+\Sigma^2)\half S_\ast$, which implies that $\Sigma^2=-1$
and $V$ does not occur, 
and which then agrees with the 
entry for $\sp(2)$ once we identify $S_\ast$ of $\so(5)$
with $F$ of $\sp(2)$.

\section{Constraints from F-theory}
\label{sec:constraints}

We now consider how global symmetries may be realized in F-theory.
As previously discussed, if we have an F-theory model defined on a
neighborhood of a contractible collection of curves, there may be additional
non-compact curves meeting the contractible collection over which
the elliptic curve degenerates further.  If there were a larger open
set containing a compactification of one of those curves, then the
Kodaria--Tate classification would have determined a gauge group for
that curve.  In the limit where the curve becomes non-compact, the
gauge coupling goes to zero and we see a global symmetry rather than
a gauge symmetry in our local model.  These are the F-theory global
symmetries we wish to compute. In all cases, we will verify
our {\em basic propostion:}\/ each maximal group from an F-theory
construction is a subgroup of the field-theoretic group determined
in section~\ref{sec:predictions}.  In most cases, we will find that
F-theory can realize the field-theoretic group.

Note that in many cases we cannot determine from the local analysis whether
the Kodaira fibers over the non-compact curve have monodromy or not, so
we assume the maximal global symmetry group compatible with the data (i.e.,
the one without monodromy).  In a few cases, though, the nature of the
intersection constrains the non-compact curve to have monodromy, thereby
reducing the size of the global symmetry group.

This is a good moment to point out a small assumption we have been forced
to make in our analysis.  The discussion of Tate's algorithm in 
\cite{newTate}
and \cite{matter1} is incomplete for Kodaira types I${}_n$, $7\le n\le9$,
and it is therefore conceivable that our analysis misses some cases
associated with those Kodaira types.

Let us consider an irreducible effective divisor $\Sigma=\{z=0\}\subseteq \{\Delta=0\}$ with self-intersection number $\Sigma\cdot\Sigma=-m$.
We introduce the condensed notation $(a,b,d)_{\Sigma}$ to indicate the orders of vanishing of $f$, $g$ and $\Delta$ along $\Sigma$, respectively.
$\Sigma$ must be topologically a $\P^1$ and its genus is $g=0$, thus we have
(as in \eqref{eq:KSigma})
\begin{align}
\label{eq:genusrat}
K_B\cdot\Sigma =-2+m~.
\end{align}
The quantities 
\begin{align}
&\ft\equiv {f \over z^a}~,  &\gt &\equiv {g \over z^b}~,    &\Deltat&\equiv{\Delta \over z^d}~,
\end{align} 
are sections of the following line bundles
\begin{align}
\label{eq:fgDeltares}
\ft& \leftrightarrow \cO(-4K_B-a\Sigma)~,		&\gt&\leftrightarrow \cO(-6K_B-b\Sigma)~, 	&\Deltat&\leftrightarrow \cO(-12 K_B-d\Sigma)~.
\end{align}
We define the {\it residual vanishings on $\Sigma$} as
\begin{align}
\label{eq:resvan}
& \at_\Sigma \equiv (-4K_B -a \Sigma)\cdot \Sigma = -4 (m-2) + ma  ~,\nonumber\\
& \bt_\Sigma \equiv (-6K_B -b \Sigma)\cdot \Sigma = -6(m-2)+mb  ~,\nonumber\\
& \dt_\Sigma \equiv (-12K_B -d \Sigma)\cdot \Sigma = -12(m-2)+md  ~.
\end{align}
These values, which are required to be non-negative since $\ft$, $\gt$, $\Deltat$ do not contain $\Sigma$ as a component, 
count the number of zeros (with multiplicity) of $\ft$, $\gt$ and $\Deltat$, respectively, when restricted to $\Sigma$.
We will refer to $\Deltat$ as the {\em residual discriminant.}  
Extending the notation above, we will indicate the triple $\eqref{eq:resvan}$ as $(\at,\bt,\dt)_\Sigma$.
Our analysis aims to determine which collections of local configurations are globally compatible with \eqref{eq:resvan}.

The first step towards our goal is to tackle the intersection between two curves.
Let $\Sigma=\{z=0\}$ and $\Sigma'=\{\sigma=0\}$ be two such curves intersecting at a point $P\equiv\Sigma\cap \Sigma'$.
We describe this situation locally by a Weierstrass model, {\it i.e.,}~by specifying the quantities in \eqref{eq:weierstr}.  
Following the discussion above we demand that the multiplicities of $f$ and $g$ at $P$ do not exceed 4 and 6, respectively.\footnote{If that is not the case, {\it i.e.}, if the multiplicity of $f$ at $P$ is at least 4 and the multiplicity of $g$ at $P$ is at least 6, it follows that the multiplicity of $\Delta$ at $P$ is at least 12.}
This basic requirement yields a first set of constraints.  In fact, we can exclude all pairwise intersections that satisfy 
\begin{align}
\label{eq:4612}
 a_\Sigma + a_{\Sigma'}&\geq4~, 		&b_\Sigma + b_{\Sigma'}&\geq6~.
\end{align}
As an example of this, we see from table \ref{t:singtypes} that a curve carrying singularity type II${}^\ast$ is in principle
only allowed to have
intersections with curves carrying singularity type I${}_n$.  These remaining cases are fully treated in appendix \ref{app:Intate}.
In general, the situation is a bit more complicated, as
\begin{align}
\ord_P f &\geq a_\Sigma + a_{\Sigma'}~, 		&\ord_P g &\geq b_\Sigma + b_{\Sigma'}~,	&\ord_P \Delta \geq d_\Sigma + d_{\Sigma'} ~.
\end{align}
It is worth noting that there are cases where the strict inequalities hold; that is, we find additional contributions to the multiplicity at $P$
beyond those coming from the degrees of vanishing along the two curves.  Geometrically, this means that our local model describes
a situation in which other components of the quantities $f$, $g$ and $\Delta$ intersect at $P$.
Hence, our criterion to discard pairwise intersections now reads
\begin{align}
\label{eq:4612bis}
\ord_Pf&\geq4~, 		&\ord_Pg&\geq6~.
\end{align}
For each intersection, it is then natural to define
\begin{align}
\label{eq:defordP}
\at_P &\equiv \ord_P \ft\big|_{z=0}~,  &\bt_P &\equiv \ord_P \gt\big|_{z=0}~,  &\dt_P &\equiv \ord_P \Deltat\big|_{z=0}~,
\end{align}
and we will use the notation $(\at_P,\bt_P,\dt_P)_{\Sigma}$. 

The second step in our analysis is to determine how to glue these local models for pairwise intersections into globally well-defined configurations.  
Let $\Sigma=\{z=0\}$ be a curve as above, 
and let $\Sigma_k$, $k=1,\dots, N$, be a collection of curves, each transversely intersecting pairwise with $\Sigma$ at the points $P_k$.
We implement the global constraints on the assembly of the local configurations; these read
\begin{align}
\label{eq:globconstrs}
\at_\Sigma &\geq \sum_{k} \at_{P_k}~, &\bt_\Sigma &\geq \sum_{k} \bt_{P_k}~, &\dt_\Sigma &\geq \sum_{k} \dt_{P_k}~.
\end{align}

Before we proceed to a more detailed case by case investigation, we wish to make a couple of general remarks.  
We say that $\Sigma=\{z=0\}$ carries {\it odd} type whenever the discriminant has the form
\begin{align}
{\Delta \over z^d} &= \left( 4 \ft^3  + 27z^p \gt^2 \right)~,
\end{align}
for some $p>0$ and $z \nmid \ft$.  We indicate this as $(a,b+B,d)_\Sigma$, where $B=0,1,\dots$.  
Setting $z=0$, the second term in the RHS vanishes identically and we find that $\dt_P = 3 \at_P$.
Similarly, we say $\Sigma$ carries {\it even} type when $(a+A,b,d)_\Sigma$, $A=0,1,\dots$, and the residual discriminant has the form
\begin{align}
{\Delta \over z^d} &= \left( 4 z^p \ft^3  + 27 \gt^2 \right)~,
\end{align}
for some $p>0$ and $z \nmid \gt$.  In this case we have instead $\dt_P = 2\bt_P$. 
 We describe the remaining cases, {\it i.e.}, I${}_n$ and I${}_n^\ast$, as {\it hybrid} types, since both $\ft\big|_{z=0}$ and $\gt\big|_{z=0}$ in principle
contribute to the residual discriminant.

In the even and odd cases, we are able to treat the general cases at once, {\it i.e.,} for any value of $A$ and $B$. 
This is done by relaxing the first condition in \eqref{eq:globconstrs} for the even case and the second in the odd case.
For hybrid curves, this is however not possible, though we will argue in the relevant case (type I${}_0^\ast$) that the maximal group is obtained
for $A=B=0$.

\subsection{Type I${}_n$}

Let $\Sigma=\{z=0\}$ carry type I${}_n$ for $n\geq2$ with self-intersection number $\Sigma^2=-m$.  The residual vanishings on $\Sigma$ are given by 
\begin{align}
(\at,\bt,\dt)_\Sigma=(8-4m, 12-6m, 24+(n-12)m)~, 
\end{align}
where $m=1,2$.  From the analysis in \cite{newTate} and \cite{matter1} 
(reviewed in appendix \ref{app:Intate}) we have
\begin{align}
\label{eq:fgz0In}
f\big|_{z=0} &= -\ff1{48}\phi_n^2~, 			&g\big|_{z=0}&=\ff1{864}\phi_n^3~, 
\end{align}
for some locally-defined function $\phi_n$, which depends on $n$.   
Except when $n=2,$ where there is no monodromy issue, we can 
additionally write\footnote{Of course, $\mu$ and $\phi_0$ depend on $n$, but we will not keep track of this index.} 
$\phi_n=\mu\phi_0^2$
with $\mu$ square-free and the monodromy is determined by whether $\mu$ vanishes\footnote{Note that our previously stated criterion for monodromy
in this case asked whether $-9g/2f$, when restricted to $z=0$, is a square
or not.  Since $-9g/2f|_{z=0} = \frac14\phi_n = \frac14\mu \phi_0^2$, this is clearly 
equivalent.} somewhere along $\Sigma$.

The case $m=2$ is very simple and we consider it first.  Notice that $\at_\Sigma=\bt_\Sigma=0,$ implying that neither quantity in 
\eqref{eq:fgz0In} can vanish.  This means that $\mu$ cannot vanish anywhere along $\Sigma,$ so the case with
monodromy cannot be realized. This matches with the fact that $\Sigma^2=-1$ for $\sp(n)$ as derived in section \ref{sec:predictions}.
Hence, there is no monodromy along $\Sigma$
and the only allowed configurations are chains of type $\cC_{n_1,\dots,n_N}\equiv[\text{I}_{n_1}, \dots, \text{I}_{n_N}]$
with associated algebras $\su(n_1)\oplus\cdots\oplus\su(n_N)$.
In particular, the maximal configuration will be $\cC_{\dt_\Sigma}$, yielding the algebra $\su(\dt_\Sigma)$ where $\dt_\Sigma=2n$.
We easily see that the basic proposition is verified here, 
as the algebra from field theory is $\so(8)$ for $n=2$ and $\su(2n)$ for $n\geq3$. 

In the rest of this section, we will deal with the case $m=1$, for which $(\at,\bt)_\Sigma=(4,6)$.
We start by proving some simple but useful results for $n\geq3$, where from \eqref{eq:fgz0In} we have $\deg\phi_n=2$ and
\begin{align}
\label{eq:constrmuphi0}
\deg\mu+2\deg\phi_0 = 2~.
\end{align}
If $\Sigma$ has monodromy, then $\deg\mu>0$ and by \eqref{eq:constrmuphi0} we get $\deg\mu =2$ and $\deg\phi_0=0$.  Otherwise, $\Sigma$ has no monodromy and $\deg\mu=0$, $\deg\phi_0=1$. 

We wish to study the restrictions that this global constraint imposes on different Kodaira singularity types when present in a configuration over $\Sigma$.
Let $n=2i$ or $n=2i+1$ and let $\Sigma'=\{\sigma=0\}$ carry type I${}_{n'},$ where $n'=2j$ or $n'=2j+1$.
To determine the maximal allowed value for $n'$ we will consider for simplicity the general forms given in \eqref{eq:Ininductive}.  The divisibility conditions read
\begin{align}
\label{eq:divcondInInp}
z,\sigma&\nmid u~,	&z^i\sigma^j & | v~,		&z^{2i}\sigma^{2j}  | w~.
\end{align}
The residual discriminant for $\Sigma$ is
\begin{align}
\label{eq:resdiscrdeltah}
\left.{\Delta\over z^n}\right|_{z=0} = u^2|_{z=0} \underbrace{\left(4uw - v^2 \right)\big|_{z=0}}_{\Deltah}~,
\end{align}
and by \eqref{eq:divcondInInp} we have that $\sigma^{n'}| \Deltah;$ that is the degree of vanishing of $\Deltah$ at $\sigma=0$ is at least $n'$.  
In other words, though $u\big|_{z=0}$ could vanish at $\sigma=0$ as well (therefore contributing to $\dt_P$), it does not increase the degree of vanishing of $\Delta$
along the transverse type I${}_{n'}$ curve.  In the remainder of this section, we assume without loss of generality that $u\big|_{z=0}$ does not vanish 
at the intersection points between $\Sigma$ and curves of type I${}_{n'}$.

Now we wish to discuss the implications of a transverse intersection with a curve carrying any of the remaining Kodaira singularity types.  
If $\Sigma$ has no monodromy, we observe from table \ref{t:sumloccontr} that 
type I${}_0^\ast$ and type I${}_{n'}^\ast$ are not allowed in any configurations, while type IV is forbidden 
for $n\geq4$ and type III for $n\geq5$.  Moreover, here $\deg \phi_0=1$, thus there can be at most one curve carrying type other than I${}_{n'}$,
which will intersect $\Sigma$ at $\phi_0|_{z=0}=0$.

In the monodromy case, $\deg \mu=2$ and configurations can admit up to two curves carrying type other than I${}_{n'}$
intersecting $\Sigma$ precisely at the roots of $\mu\big|_{z=0}$. 
To summarize, the relevant configurations are chains of the type $\cC_{n_1,\dots,n_N}$ with one or two 
other singularity types attached. However, since $\cC_{n_1,\dots,n_N}$ yields the algebra $\oplus_i \su(n_i)$, when looking for the maximal
global symmetry group we will just consider $\cC_{\sum_i n_i}$ with algebra $\su(\sum_i n_i)$. 

Perhaps a comment is useful here. 
We stated above that Kodaira types other than I${}_{n'}$ can only intersect $\Sigma$ at the roots of $\phi_n|_{\Sigma}$.
There is an example that seems to contradict this. Let us consider an intersection with type IV along the locus $\{\sigma=0\}$, 
then both $f$ and $g$ vanish to order 2 in $\sigma$ and 
it seems that $\phi_n= -9g/2f|_{z=0}$ need not vanish. 
However, the point is that the quantity $g/f$ is evaluated along $\Sigma$, where $g|_{z=0}$ and $f|_{z=0}$
can have higher orders of vanishing. These are determined carefully in appendix \ref{app:Intate}, and it turns out that 
for all intersections, except with type I${}_{n'}$, $\phi_n$ does indeed vanish. 

Having summarized some of the general features of the pertinent local models, let us proceed to a detailed analysis. 

\subsection*{$\bf{n=2}$}

In this case, there is no monodromy ambiguity and
we see from table \ref{t:sumloccontr} that types III, IV, I${}_0^\ast$ and I${}_{n'}^\ast$ can be part of a configuration on $\Sigma$.  Here,
\begin{align}
{\Delta\over z^2} {\Big |}_{z=0} = \phi^2 \left( \phi\gt_2 - f_1^2\right)~,
\end{align}
where the quantities on the RHS are restricted to the locus $\{z=0\}$.  In the language of \eqref{eq:resdiscrdeltah}, we define 
\begin{align}
\label{eq:deltahn2}
\Deltah \equiv \phi\gt_2 - f_1^2~.
\end{align}
From \eqref{eq:fgz0In}, it follows $\deg\phi=2$ and there can be up to two curves carrying type other than I${}_{n'}$ in a configuration. 
Let $\Sigma'=\{\sigma=0\}$ carry a given Kodaira type and let $P\equiv \Sigma\cap\Sigma'$. Our strategy is to determine 
the lowest order of vanishing of $\Deltah$ at P, and introducing a non-standard notation, we will denote
this value by $\ord_P \Deltah$. This is easily done by recalling the general form for type I${}_2$
\begin{align}
\label{eq:I2gensol}
f&=-\frac1{48}\phi^2+f_1z+ O(z^2)~,		&g&=\frac1{864}\phi^3 -\frac1{12} \phi f_1 z+(\gt_2-\frac{1}{12}\phi f_2)z^2+O(z^3)~,
\end{align}
as well as the data from table \ref{t:singtypes}. In fact, let $\Sigma'$ carry type III, then $\sigma|f$ and $\sigma^2|g$, making $\sigma|\{\phi,f_1\}$
and $\sigma^2|\gt_2$. Plugging these into \eqref{eq:deltahn2}, we obtain $\sigma^2|\Deltah$. 
If $\Sigma'$ carries type IV, then $\sigma^2|f$ and $\sigma^2|g,$ and hence $\sigma^2|f_1$ while everything else is unchanged from the case above, 
in turn yielding $\ord_P\Deltah=3$. Next, we consider the case of $\Sigma'$ carrying type I${}_p^\ast$ for $p\geq0$. 
The argument here is somewhat different, as it suffices to realize that
\begin{align}
6+p=\ord_P \Delta = 2\ord_P \phi+\ord_P\Deltah~.
\end{align}
Now, the solution \eqref{eq:I2gensol} is equivalent to \eqref{eq:Ininductive} with the following identifications
\begin{align}
u&=\frac14\phi~,		&v_i&=f_i~,\quad i\geq1~,		&w_j&=\gt_j~,	\quad j\geq2~.
\end{align}
The Tate form for I${}_p^\ast$ (regardless of monodromy) prescribes $\sigma|u$ but $\sigma^2 \nmid u$, hence $\ord_P\phi=1$.\footnote{In other words, if $\ord_P\phi>1$ then we are in a different brach of Kodaira's classification.} This determines $\ord_P\Deltah=4+p$.
We summarize all of this in the following table
\begin{align}
\label{eq:nevendivconds}
\nonumber
\xymatrix@C=7mm@R=0mm{
				&\text{I}_0^\ast		&\text{I}_{p}^\ast		&\text{III}		&\text{IV}		&\text{I}_{n'} \\
\ord_P\Deltah		&4				&4+p				&2			&3			&n'
}
\end{align}
This also allows us to reduce the types of configurations we need to consider.
From the data above, as far as algebras are concerned 
\begin{align}
&[\dots,\text{III},\dots, \text{I}_{n'},\dots]\subset [\dots,\text{I}_{n'+2},\dots]~,
&&[\dots,\text{IV},\dots, \text{I}_{n'},\dots]\subset [\dots,\text{I}_{n'+3},\dots]~.
\end{align}
Hence, we avoid considering type III and type IV.
Finally, recalling that $\deg \Deltah=10$, we obtain the following relevant configurations
\begin{align}
\nonumber
\xymatrix@C=7mm@R=0mm{
\text{configuration(s)}						&\text{algebra(s)}\\
\text{I}_{10}								&\su(10)\\
\text{I}_p^\ast, \text{I}_{6-p}, {\ 0\leq p \leq 6}		&\so(8+2p)\oplus\su(6-p) \\
\text{I}_p^\ast, \text{I}_{q}^\ast, \text{I}_{2-p-q},\ { 0\leq \{p,q,p+q\} \leq 2} 	&\so(8+2p)\oplus\so(8+2q)\oplus\su(2-p-q) \\
}
\end{align}
We conclude that the global symmetry of
the maximal configuration [I${}_6^\ast$] agrees with the
field-theoretic prediction of  $\so(20)$.

\subsection*{$\bf{n\geq3}$ odd}

Here $\dt_\Sigma=12+n,$ and from appendix \ref{app:Intate} we have
\begin{align}
{\Delta\over z^n}\Big|_{z=0} = u_0^2 \left( 4 u_0 w_{2i+1} +4u_1w_{2i}-2v_i v_{i+1} \right)~,
\end{align}
where $u_0=\mu\phi_0^2$ and $n=2i+1$.  As before, we assume throughout 
the rest of this section the generic form \eqref{eq:Ininductive}, 
and we can rewrite the last expression as
\begin{align}
\label{eq:resdiscrnodd}
{\Delta\over z^n}\Big|_{z=0} =\frac1{16} \mu^3 \phi_0^4\left( \phi_0^2 w_{2i+1} +t_i^2 u_1-\phi_0t_i v_{i+1} \right)~,
\end{align}
denoting by $\Deltah$ the term inside the brackets. Even for the cases $n=3$ and $n=5$ the expressions above are completely general, as
there is an isomorphisms between the explicit forms \eqref{eq:I3expl} and \eqref{eq:I5} and the inductive form \eqref{eq:Ininductive}, given by (for $n=5$)
\begin{align}
u_0 &= \frac14\mu\phi_0^2~,				&u_1&=\phi_1~,
&v_2 & = \frac12 \mu\phi_0\psi_2~,			&v_k &= f_k~, \ k\geq3~,
&w_4 & = \frac14 \mu \psi_2^2 ~,			&w_l &= \gh_l~,	 \ l\geq5~.
\end{align}

First, let us suppose that there is monodromy on $\Sigma$.  From above we have that $\deg \mu =2$ and $\deg \phi_0=0$; we set $\phi_0\equiv1$ for concreteness. 
This implies $\deg \Deltah = \dt_\Sigma - 6 = 6+n$, yielding I${}_{6+n}$ as the largest
type I${}_{n'}$ allowed in any configuration. 

To take into account the presence of other singularity types we need to determine, as before, the lowest vanishing order of $\Deltah$ along 
$\Sigma'=\{\sigma=0\}$.
We recall from \eqref{eq:genforsvwnodd} the expressions 
\begin{align}
\label{eq:genforsvwnodd2}
u &= \frac14 \mu + u_1 z + O(z^2)~,\nonumber\\
v &= \frac12 \mu t_i z^i + v_{i+1}z^{i+1} + O(z^{i+2}) ~,\nonumber\\
w &= \frac14 \mu t_i^2 z^{2i} + w_{2i+1}z^{2i+1} + O(z^{2i+2}) ~.
\end{align}
When $\Sigma'$ carries type III, we have that $\sigma|\{u,v\}$ and $\sigma^2|w$. 
Since $\sigma^2$ must divide $\mu t_i^2$ and $\mu$ is square-free, it follows that $\sigma | t_i$. 
All of this implies that $\sigma^2|\Deltah$ for type III. When $\Sigma'$ carries type IV without monodromy 
(if there is monodromy, the gauge algebra is $\su(2)$ and the discussion follows the one above for type III),
we have $\sigma|u$ and $\sigma^2|\{v,w\}$.
Hence, the terms in
\begin{align}
g&=\frac2{27} u^3 - \frac13 uv + w
\end{align}
have degrees of vanishing 3, 3 and 2 along $\{\sigma=0\},$ respectively. Monodromy forces $g/\sigma^2|_{\sigma=0}= w$ to be a square, and from \eqref{eq:genforsvwnodd2}, we obtain
\begin{align}
\label{eq:wisaquare}
\frac{w}{\sigma^2}\Big|_{\sigma=0} = \wt_{2i+1}\big|_{\sigma=0} z^{2i+1} + \cdots~,
\end{align}
where $w_{2i+1} = \sigma^2 \wt_{2i+1}$. Whether \eqref{eq:wisaquare} is a square or not is contingent upon whether $\wt_{2i+1}$ vanishes identically or it is further divisible by $\sigma$.
In either case, it is straightforward to verify that $\ord_P \Deltah = 3$.

In the case $\Sigma'$ supports type I${}_0^\ast$, there are two cases we need to distinguish. In fact, while $\sigma|u$ and $\sigma^2|v$ always hold, 
we have that $\sigma^3|w$ for I${}_0^\ast{}^{\text{ns}}$ (algebra $\g_2$) and $\sigma^4|w$ for both I${}_0^\ast{}^{\text{ss}}$ and I${}_0^\ast{}^{\text{s}}$ 
(algebras $\so(7)$ and $\so(8),$ respectively). For the latter case, we will consider only I${}_0^\ast{}^{\text{s}}$ as it leads to the largest global symmetry.
It follows that $\sigma|t_i$ for I${}_0^\ast{}^{\text{ns}}$ and $\sigma^2|t_i$ for I${}_0^\ast{}^{\text{s}}$, and \eqref{eq:resdiscrnodd} 
determines $\sigma^3| \Deltah $ for I${}_0^\ast{}^{\text{ns}}$ and 
$\sigma^4| \Deltah$ for I${}_0^\ast{}^{\text{s}}$.

The remaining case we need to analyze is when $\Sigma'$ carries type I${}_p^\ast$ for $p\geq1$. First suppose that $p=2n'+1$; then $\sigma|u$, 
$\sigma^{n'+3}|v$ and $\sigma^{2n'+4}|w$, which in turn implies $\sigma^{n'+2}|t_i$. This leads to $\ord_P \Deltah = 2n'+4=p+3$, and this holds
for I${}_p^\ast{}^{\text{ns}}$. 
The condition for monodromy is whether
\begin{align}
\frac{w}{\sigma^{2n'+4}}\Big|_{\sigma=0} = \wt_{2i+1}\big|_{\sigma=0} z^{2i+1} + \cdots~,
\end{align}
where $w_{2i+1}=\sigma^{2n'+4}\wt$, has a square root. As before, in order to have no monodromy $\wt_{2i+1}|_{\sigma=0}=0,$ and this
makes $\ord_P \Deltah=2n'+5=p+4$. The even case, $p=2n'$, is similar, but we repeat the argument for completeness. Here we have $\sigma|u$, 
$\sigma^{n'+2}|v$ and $\sigma^{2n'+3}|w$, which leads to $\sigma^{n'+1}|t_i$. 
It is convenient to define $\mu=\sigma \tilde\mu$, $u_1=\sigma \ut_1$, 
$t_i = \sigma^{n'+1}\tilde t_i$, $v_{i+1}=\sigma^{n'+2}\vt_{i+1}$ and $w_{2i+1}=\sigma^{2n'+3}\wt_{2i+1}$.
This gives $\ord_P \Deltah = 2n'+3=p+3$, valid for I${}_p^\ast{}^{\text{ns}}$.
In order to have no monodromy, the quantity
\begin{align}
\label{eq:monodrcondevenp}
4 \left(\frac{u}{\sigma}\right)\Big|_{\sigma=0} \left(\frac{w}{\sigma^{2n'+3}}\right)\Big|_{\sigma=0} - \frac{v}{\sigma^{2n'+2}} \Big|_{\sigma=0} = 
\tilde\mu \left( {\tilde{t}_i}^2 \ut_1 + \wt_{2i+1} - \tilde t_i  \vt_{i+1}\right)\big|_{\sigma=0} z^{2i+1} + O(z^{2i+2})~,
\end{align}
must be a square. Therefore, the coefficient of the first term on the RHS of \eqref{eq:monodrcondevenp} must vanish 
and in particular it coincides, up to an irrelevant factor of $\tilde\mu$, with \eqref{eq:resdiscrnodd},  giving $\ord_P \Deltah = p+4$.
We summarize these results in the following table
\begin{align}
\label{eq:nodddivcondsbis}
\nonumber
\xymatrix@C=7mm@R=0mm{
				&\text{I}_0^\ast{}^{\text{ns}}		&\text{I}_0^\ast{}^{\text{s}}		&\text{I}_{p}^\ast{}^{\text{ns}}		&\text{I}_{p}^\ast{}^{\text{s}}		&\text{IV}		&\text{III}		&\text{I}_{n'} \\
\ord_P\Deltah		&3						&4							&3+p						&4+p						&3			&2			&n'
}
\end{align}
The argument we provided in the $n=2$ case for type III and type IV goes through here as well, since for these, the values of $\ord_P\Deltah$ are identical. 
Thus, we will exclude configurations containing them. 
The remaining relevant configurations are
\begin{align}
\nonumber
\xymatrix@C=7mm@R=0mm{
\text{configuration(s)}															&\text{algebra(s)}\\
\text{I}_{n+6}																	&\su(n+6)\\
\text{I}_p^\ast{}^{\text{s}}, \text{I}_{n+2-p}, \ \ 0\leq p \leq n+2								&\so(8+2p)\oplus\su(n+2-p) \\
\text{I}_p^\ast{}^{\text{ns}}, \text{I}_{n+3-p}, \ \ 0\leq p \leq n+3								&\so(7+2p)\oplus\su(n+3-p) \\
\text{I}_p^\ast{}^{\text{s}}, \text{I}_{q}^\ast{}^{\text{s}}, \text{I}_{n-2-p-q},\ { 0\leq \{p,q, p+q\} \leq n-2}	&\so(8+2p)\oplus\so(8+2q)\oplus\su(n-2-p-q)\\
\text{I}_p^\ast{}^{\text{s}}, \text{I}_{q}^\ast{}^{\text{ns}}, \text{I}_{n-1-p-q},\ { 0\leq \{p,q, p+q\} \leq n-1}	&\so(8+2p)\oplus\so(7+2q)\oplus\su(n-1-p-q)\\
\text{I}_p^\ast{}^{\text{ns}}, \text{I}_{q}^\ast{}^{\text{ns}}, \text{I}_{n-p-q},\ { 0\leq \{p,q, p+q\} \leq n}	&\so(7+2p)\oplus\so(7+2q)\oplus\su(n-p-q)
}
\end{align}
By inspection, we see that within each row of the above table, except the first, the algebra is maximal when $p$ or $p+q$ assume the highest allowed value, 
{\it i.e.}, when the last curve is I${}_0$.
This also tells us that $[\text{I}_p^\ast{}^{\text{s}}, \text{I}^\ast{}^{\text{s}}_{n-2-p}] \subset [\text{I}^\ast{}^{\text{s}}_{n+2}]$
and $[\text{I}_p^\ast{}^{\text{s}}, \text{I}^\ast{}^{\text{ns}}_{n-1-p}]\subset [\text{I}^\ast{}^{\text{ns}}_{n+3}]$ and of course 
$[\text{I}_{n+6}]\subset [\text{I}^\ast{}^{\text{s}}_{n+2}]\subset [\text{I}^\ast{}^{\text{ns}}_{n+3}]$. The locally maximal configurations are then given by
\begin{align}
\nonumber
\xymatrix@C=7mm@R=0mm{
\text{configuration(s)}															&\text{algebra(s)}\\
\text{I}^\ast{}^{\text{ns}}_{n+3}						&\so(13+2n) \\
\text{I}_p^\ast{}^{\text{ns}}, \text{I}^\ast{}^{\text{ns}}_{n-p}, \ { 0\leq p \leq \ff{n+1}2}			&\so(7+2p)\oplus\so(7+2n-2p)
}
\end{align}
The predicted algebra is $\so(16+2(n-1))$.  It is easy to check
that the basic proposition holds for each value of $n$.
This example shows two important characteristics of our findings, as we mentioned in the discussion above.  First, we 
verified explicitly that the algebra predicted through field theory
 is not always realizable in F-theory. 
Second, this model exhibits multiple relatively maximal algebras, each subalgebras of the maximal algebra from field theory.

If there is no monodromy on $\Sigma$, we set $\mu\equiv1$ and $\deg\phi_0=1$.
Then $\deg \Deltah = \dt_\Sigma-4=8+n$.
We recall that type I${}_0^\ast$ and type I${}_n^\ast$ are forbidden in all configurations. 
However, as mentioned above, for $n=3$ type III and type IV are still allowed, so we obtain
\begin{align}
\xymatrix@C=7mm@R=0mm{
\text{configuration}							&\text{algebra}\\
\text{I}_{11}								&\su(11)\\
\text{IV}, \text{I}_8							&\su(3)\oplus\su(8) \\
\text{III}, \text{I}_9							&\su(2)\oplus\su(9) \\
}
\end{align}
These are all subalgebras of $\su(11)$ (which is in turn a
subalgebra of field-theoretic prediction $\su(12)$).  For $n\geq5,$ the only relevant configurations are of the form
$\cC_{n_1,\dots,n_N}$, and the maximal algebra among these results from $\cC_{\dt-4}$.  This yields $\su(8+n)$, which coincides
 with the prediction from field theory.

\subsection*{$\bf{n\geq4}$ even}

In this case, we set $n=2i$ and the residual discriminant assumes the form 
\begin{align}
\label{eq:nevenresdiscr}
{\Delta\over z^n} \Big|_{z=0} = u_0^2 \left( 4 u_0 w_{2i}-v_i^2 \right)~,
\end{align}
where again $u_0=\ff14\mu\phi_0^2$, and 
\begin{align}
\label{eq:genforsvwneven2}
u &= \frac14 \mu\phi_0^2 + u_1 z + \dots~,
&v &= v_iz^i + v_{i+1}z^{i+1} + \cdots ~,
&w &=w_{2i} z^{2i} + w_{2i+1}z^{2i+1} + \cdots ~.
\end{align}
We define $\Deltah \equiv 4 u_0 w_{2i}-v_i^2$.
 
Suppose first that $\Sigma$ has monodromy. In this case the form \eqref{eq:nevenresdiscr} is generic within the assumptions of this paper. In fact,
we have already showed that the inductive form \eqref{eq:Ininductive} reproduces the generic form for type I${}_5$ regardless of monodromy, and 
in particular for type I${}_4$. For type I${}_6$ this is not true in general, but it holds when we impose that $\Sigma$ has monodromy. In this case, 
recall that $\deg\mu=2$ and $\deg\phi_0=\deg\alpha+\deg\beta=0$ and we set wlog $\alpha=\beta\equiv1$. The identification 
between the two sets of local functions is
\begin{align}
u_0 &= \frac14\mu~,						&u_1&=\nu~,					&u_2&=\phi_2~,\nonumber\\
v_3 & = \frac13 \nu\phi_2 - 3\lambda~,		&v_4 &= f_4 + \frac13\phi_2^2~,		&v_k &= f_k~, \ k\geq5~,\nonumber\\
w_6 & = \gh_6 + \frac13 f_4 \phi_2 + \frac1{27}\phi_2^3~,		&w_l &= \gh_l~,	 \ l\geq7~.
\end{align}
Hence, the only case which is not captured in full generality \footnote{As stated above, we are making an assumption for $n=7,8,9$.} by \eqref{eq:nevenresdiscr} is
$n=6$ with no monodromy, which we will discuss separately. 

Let us proceed and determine the lowest order of vanishing of $\Deltah$ for intersections with 
different Kodaira types. Given the simpler form of the solution \eqref{eq:genforsvwneven2}
with respect to \eqref{eq:genforsvwnodd2}, the analysis will be less involved. 
Let $\Sigma'=\{\sigma=0\}$ carry type III, then $\sigma|\{u,v\}$, $\sigma^2|w$ and $\sigma^2|\Deltah$. When $\Sigma'$ carries type IV, the only
modification is that now $\sigma^2|v$, yielding $\sigma^3|\Deltah$. Now let $\Sigma'$ carry type I${}_p^\ast$, where $p\geq0$. 
In what follows, we do not need
to distinguish between the different monodromy cases, so we will just assume the largest induced algebra. If $p=2n'+1$ we have 
$\sigma|u$, $\sigma^{n'+3}|v$, $\sigma^{2n'+4}|w$ and $\sigma^{2n'+5}|\Deltah$; if $p=2n'$ we have 
$\sigma|u$, $\sigma^{n'+2}|v$, $\sigma^{2n'+3}|w$ and $\sigma^{2n'+4}|\Deltah$. 
We summarize these results in the following table
\begin{align}
\label{eq:nevendivconds2}
\xymatrix@C=7mm@R=0mm{
				&\text{I}_0^\ast		&\text{I}_{p}^\ast		&\text{IV}		&\text{III}		&\text{I}_{n'} \\
\ord_P\Deltah		&4				&4+p				&3			&2			&n'
}
\end{align}
As before, we do not need to consider configurations involving type III and type IV.
Since $\deg\Deltah=\dt_\Sigma-4=8+n$, we obtain 
\begin{align}
\nonumber
\xymatrix@C=7mm@R=0mm{
\text{configuration(s)}											&\text{algebra(s)}\\
\text{I}_{8+n}													&\su(8+n)\\
\text{I}_p^\ast, \text{I}_{4+n-p}, \ \ 0\leq p \leq n+4						&\so(8+2p)\oplus\su(4+n-p) \\
\text{I}_p^\ast, \text{I}_q^\ast, \text{I}_{n-p-q}, \ \ 0\leq \{p,q,p+q\} \leq n		&\so(8+2p)\oplus\so(8+2q)\oplus\su(n-p-q) \\
}
\end{align}
We see that there is a maximal algebra, namely $\so(16+2n)$, given by $[\text{I}_{n+4}^\ast]$.
This coincides with the prediction for each $n$.

If there is no monodromy on $\Sigma$ (and $n\neq6$), the only other singularity type other that I${}_{n'}$ is type III for $n=4$, for which 
$\dt_P\geq4$, as can be read from \eqref{eq:nevendivconds2} and \eqref{eq:nevenresdiscr}. 
The relevant configurations in this case are therefore
\begin{align}
\nonumber
\xymatrix@C=7mm@R=0mm{
\text{configuration}							&\text{algebra}\\
\text{I}_{12}								&\su(12)\\
\text{III}, \text{I}_{10}							&\su(2)\oplus\su(10)
}
\end{align}
The prediction from field theory is $\su(12)$, hence we have verified our 
basic proposition.   If $n>4,$
the analysis is similar to the odd case: the only relevant configurations are $\cC_{n_1,\dots,n_N}$ and the maximal is $\cC_{\dt-4}$.
This again yields $\su(8+n)$, matching the field theory result.

The remaining case to be analyzed is $n=6$ without monodromy. From \eqref{eq:I4} we have
\begin{align}
\label{eq:ftgtsu6}
\ft\big|_{z=0} & = -\ff1{48}\alpha^4\beta^4 ~, 			&\gt\big|_{z=0} & = \ff1{864}\alpha^6\beta^6~,
\end{align}
as well as
\begin{align}
 \Deltat\big|_{z=0} &= \ff1{432}\alpha^4\beta^3 \underbrace{\left( 27\alpha^2\beta^3\gh_6+9\alpha^2\beta^2\phi_2f_4 +\alpha^2\phi_2^3
  -243\lambda^2 \beta^3+ 54\phi_2\nu\lambda\beta^2 - 3\beta\nu^2\phi_2^2\right)}_{\Deltah} .
\end{align}
Now, since $\Sigma^2=-1$ we have that $(\at,\bt)_{\Sigma}=(4,6)$, which together with \eqref{eq:ftgtsu6} implies 
\begin{align}
\deg \alpha + \deg\beta =1~.
\end{align}
Also, we have that $\dt_\Sigma = 18$ and
\begin{align}
\label{eq:dsigmaalbe}
\dt_\Sigma  = 4\deg \alpha +3 \deg \beta + \deg \Deltah~.
\end{align}
We recall that the relevant configuration $\cC_{n_1,\dots,n_N}$ only has  support on $\Deltah$, that is $\sigma_i\nmid \{\alpha,\beta\}$, where $\{\sigma_i=0\}$ supports
the i-th curve in the configuration. We can then rewrite \eqref{eq:dsigmaalbe} as
\begin{align}
18 - \sum_in_i   \geq 4\deg \alpha +3 \deg \beta~.
\end{align}
There are two possibilities:
\begin{enumerate}
\item $\deg \alpha=1,\ \deg\beta=0$. Then $\sum_in_i   \leq 14$ and the maximal configuration is $\cC_{\dt_\Sigma-4}$ with algebra $\su(14)$.
This is the algebra associated to the $\su(6)$ row of table \ref{t:fieldthypreds}.
\item $\deg \alpha=0,\ \deg\beta=1$. Here, we instead have $\sum_in_i   \leq 15$ and the maximal configuration is $\cC_{\dt_\Sigma-3}$ with algebra $\su(15)$.
This corresponds to the alternative possibility $\su(6)^\ast$.
\end{enumerate}

\subsection{Type III}

Let $\Sigma=\{z=0\}$ carry singularity type III and $\Sigma\cdot\Sigma=-m$.
Plugging $(a,b,d)_\Sigma=(1,2+B,3)_\Sigma$ in \eqref{eq:resvan} we compute $(\at,\bt,\dt)_\Sigma = (8-3m,12-4m+B m,24-9m)$.
We now proceed to separately analyze the two cases $m=1,2$. 

Let us start with $m=2$. First of all, we need to determine which curves locally admit allowed intersections. 
From \eqref{eq:4612} and appendix \ref{app:Instartate} we see that type IV${}^\ast$, III${}^\ast$ and II${}^\ast$ are forbidden, 
as well as type I${}_n^\ast$ for $n\geq1$.
From table \ref{t:sumloccontr}, we see that an intersection with I${}_n$ is possible for each $n$, which together with type III, IV and I${}_0^\ast$
constitute the blocks for building configurations.  As we remarked above, we do not need to worry about $\bt_P$ for odd types.  
Thus, we list the above singularities types with their contributions to $\at_P$, borrowing the appropriate entries from table \ref{t:singtypes} and table \ref{t:sumloccontr}
\begin{align}
\nonumber
\xymatrix@C=7mm@R=0mm{
		&\text{III} 		&\text{IV} 		&\text{I}_0^\ast		&\text{I}_n\\
\at_P		&1			&2			&\geq2			&\lceil n/2\rceil
}
\end{align}
Now, since $\at_\Sigma=2$ we must exclude I${}_{n\geq5}$, and the various possibilities compatible with \eqref{eq:globconstrs} are
\begin{align}
\nonumber
\xymatrix@C=10mm@R=0mm{
\text{configuration}		&\text{algebra}	\\
\text{III, III}			&\su(2)\oplus\su(2)	\\
\text{I}_2,\ \text{III}		&\su(2)\oplus\su(2)	\\
\text{I}_2,\ \text{I}_2		& \su(2)\oplus\su(2)\\
\text{I}_3				& \su(3) \\
\text{I}_4 				& \su(4) \\	
\text{IV}				& \su(3) \\
\text{I}_0^\ast			& \so(7) \text{ or } \mathfrak{g}_2
}
\end{align}
Note that we cannot assume that
an I${}_0^*$ meeting a type III supports an $\so(8)$
algebra because that would require all intersection points with
the residual discriminant (even the ones not near the current intersection)
to have even multiplicity.  And indeed, the analysis of
\cite{Ohmori:2015fk} shows that it can be at most $\so(7)$. 
Explicitely, let $\Sigma'=\{\sigma=0\}$ carry type I${}_0^\ast$, then
\begin{align}
f&=z\sigma^2 f_0~, 		&g&= z^2\sigma^3 g_0(\sigma)~,
\end{align}
where $f_0$ is a non-zero constant and $\deg g_0 (\sigma) = 2B+1$ such that $g_0(0)\neq0$. 
The residual discriminant for $\Sigma'$ reads 
\begin{align}
{\Delta \over \sigma^6}\Big|_{\sigma=0} = z^3 \left( 4 f_0^3 + 27 z g_0^2(0)\right)~,
\end{align}
which is not a square in $z$ and the algebra supported on $\Sigma'$ cannot be $\so(8)$.
We conclude that there is an algebra, namely $\so(7)$, that admits as a subalgebra every other entry in the list and is the desired maximal global
symmetry algebra in this case.

Let us turn to the case $m=1$.
The local models for the intersections are the same as above, but this time we have $\at_\Sigma=5$.  This leads to a larger set of possibilities for the 
intersecting curves, which we do not reproduce here.  The result is that there are three different models that exhibit maximal global symmetry:
\begin{align}
\nonumber
\xymatrix@C=10mm@R=0mm{
\text{configuration}							&\text{algebra}	\\
\text{I}_0^\ast,\ \text{I}_0^\ast,\ \text{I}_2			& \so(7)\oplus\so(7)\oplus\su(2)		\\
\text{I}_0^\ast,\ \text{I}_6						& \so(7)\oplus\su(6)\\
\text{I}_{10}								& \su(10) 
}
\end{align}
As before, we have indicated the monodromy choices in each case making the global symmetry algebra relatively maximal and we used
the fact that type I${}_0^\ast$ cannot support an $\so(8)$ algebra.  
We notice that these algebras are all contained in the
algebra $\so(20)$ predicted from field theory.

\subsection{Type IV}

We now turn to the analysis of a Kodaira type IV curve along $\Sigma=\{z=0\}$ with $(a,b,d)_\Sigma=(2+A,2,4)$, 
for which the residual vanishings are given by
$(\at,\bt,\dt)_\Sigma=(8-2m+A m,12-4m,24-8m)$.
The possibilities for the self-intersection are $m=1,2,3$.  Moreover, there are two possibilities for the monodromy on a type IV,
the condition being whether $\gt\big|_{z=0}$ is a square.  
If $m=3$, the gauge algebra must be $\su(3)$, and there is no matter
(and hence no global symmetry).

For $m=2,$ we have $\bt_\Sigma =4$.  Let us start by listing configurations that satisfy \eqref{eq:globconstrs}.
These are
\begin{align}
\nonumber
\xymatrix@C=10mm@R=0mm{
\text{configuration}					&\text{algebra}	\\
\text{I}_0^\ast						& \so(7) \text{ or } \mathfrak{g}_2\\
\text{III, III}						& \su(2)\oplus\su(2) \\
\text{I}_3,\ \text{I}_3					& \su(3)\oplus\su(3)\\
\text{IV, IV}						& \su(3)\oplus\su(3) \\
\text{IV},\ \text{I}_3					& \su(3)\oplus\su(3) \\
\text{I}_4							& \sp(2) 
}
\end{align}
Now we will determine how the monodromy will impose different restrictions in each case.
\begin{itemize}
\item \underline{$\su(2)$.} In this case, $\gt\big|_{z=0}$ is not allowed to have all double roots. 
Therefore, we discard the configurations for which each intersection yields an even value for $\bt_P$.  
All configurations in the list above except the first fall into this category.
As in the case of type III,
the type I${}_0^\ast$ gives $\so(7)$ (or possibly $\mathfrak{g}_2$)
as global algebra and it is maximal in this case.
\item \underline{$\su(3)$.} In this case, $\gt\big|_{z=0}$ is a square and hence all its roots are double roots.  Moreover, $\Sigma$ is not allowed 
to intersect $\Sigma'=\{\sigma=0\}$ carrying type I${}_0^\ast$ since $\bt_P\geq3$, where $P\equiv\Sigma\cap\Sigma'$, 
is increased at least to $\bt_P=4$.  Thus, at $P$ we are in the case \eqref{eq:4612}
and the intersection is non-minimal.  The remaining configurations are in principle allowed and yield
$\su(3)\oplus\su(3)$ and $\sp(2)$ as maximal global symmetries, which indeed are both subalgebras of the algebra $\su(6)$ predicted from field theory.
\end{itemize}

If $m=1$, then $\bt_\Sigma=8$ and the list of possible configurations is
\begin{align}
\nonumber
\xymatrix@C=10mm@R=0mm{
\text{configuration}					&\text{algebra}	\\
\text{I}_0^\ast,\ \text{I}_0^\ast,\ \text{IV}	& \so(7)^{\oplus2}\oplus\su(3)\\
\text{I}_0^\ast,\ \text{I}_0^\ast,\ \text{I}_3	& \so(7)^{\oplus2}\oplus\su(3)\\
\text{I}_0^\ast,\ \text{IV, IV}			& \so(7)\oplus\su(3)^{\oplus2}\\
\text{I}_0^\ast,\ \text{I}_5				& \so(7)\oplus\sp(2)\\
\text{IV, IV, IV, IV}					& \su(3)^{\oplus4} \\
\text{IV},\ \text{I}_6					& \su(3)\oplus\sp(3) \\
\text{I}_8							& \sp(4) \\
\text{I}_4,\ \text{IV, IV}				& \sp(2)\oplus\su(3)^{\oplus2}
}
\end{align}
Again, we discuss the two monodromy cases separately.
\begin{itemize}
\item \underline{$\su(2)$.} Let us start by showing that the configuration [I${}_0^\ast$, I${}_0^\ast$, IV] leads to a global symmetry that is not larger than
$\so(7)\oplus\so(7)\oplus\su(3).$  In fact, let $\{\sigma_1=0\}$, $\{\sigma_2=0\}$ and $\{\sigma_3=0\}$ be the curves carrying Kodaira type 
IV and the two type I${}_0^\ast,$ respectively.  Then we have the following divisibility conditions
\begin{align}
f&=z^2\sigma_1^2\sigma_2^2\sigma_3^2 f_0(\sigma)~, 		&g&=z^2\sigma_1^2\sigma_2^3\sigma_3^3 g_0~, 
\end{align}
where $g_0$ is a non-zero constant and $\deg f_0(\sigma)=\alpha$ and is such that $f_0(\sigma_i=0)\neq0$ for $i=1,2,3$.
Looking at $g/\sigma_1^2|_{\sigma_1=0},$ we see it is a perfect square in $z$, thus the gauge algebra along the intersecting IV
is indeed $\su(3)$ (this holds since the set-up already fills out $\bt_\Sigma$).
Now, to determine the monodromy on the I${}_0^\ast$ curves, we look at the residual discriminant restricted to 
one of these curves, say $\{\sigma_2=0\}$, which reads
\begin{align}
{\Delta\over\sigma_2^6}\Big|_{\sigma_2=0} = z^4 \left( 4z^2\sigma_1^6\sigma_3^6 f_0^3+27\sigma_1^4\sigma_3^6 g_0^2\right)\big|_{\sigma_2=0}~,
\end{align}
which is not a perfect square in $z$.  Thus, the algebra cannot be $\so(8)$ for either of the two curves carrying type I${}_0^\ast$.  
We are therefore restricted to
at most $\so(7)\oplus\so(7)\oplus\su(3)$.  This argument actually applies for any 
configuration that contains type I${}_0^\ast$, in particular for the first four configurations of the above list.  
Finally, as before, we discard configurations in which each intersection yields an even value for $\bt_P$, for example a configuration of four type IV.  
This leaves $\so(7)\oplus\so(7)\oplus\su(3)$ as the maximal algebra.

\item \underline{$\su(3)$.} Here we recall that a configuration containing type I${}_0^\ast$ is not allowed.  
Therefore, we are left with the last four configurations of the list above, and all of these lead to subalgebras of the field-theoretic prediction  $\su(12)$.
\end{itemize}

\subsection{Type I${}_0^\ast$}

Let us turn our attention to $\Sigma=\{z=0\}$ carrying type I${}_0^\ast$ with $\Sigma^2=-m$ and $(2+A,3+B,6)_{\Sigma}$, where
$A$ and $B$ cannot be both nonzero.  
The residual vanishings on $\Sigma$ are $(\at,\bt,\dt)_\Sigma=(8-2m+A m,12-3m+B m,24-6m)$.
The local analysis derived in appendix \ref{app:Intate} still applies, but it must 
be supplemented by a more careful treatment of the global constraints to be imposed on the Weierstrass model.  
We are going to prove some simple results 
about the {\it global} models involving intersections between type I${}_0^\ast$ and various curves carrying type I${}_n$, for some $n$.  
In particular, we look for upper bounds on $n$ such that type I${}_n$ is allowed to intersect type I${}_0^\ast$, 
although {\it locally} the intersection is consistent for any $n$ as can be seen from table \ref{t:sumloccontr}.  

There are three possibilities for the gauge algebra on Kodaira type I${}_0^\ast$, determined by the behavior of the monodromy cover
\begin{align}
\label{eq:moncovI0st}
\psi^3 + {f\over z^2}\Big|_{z=0} \psi + {g\over z^3}\Big|_{z=0}~.
\end{align}
If the monodromy cover is irreducible, the gauge algebra is $\g_2$; if it has two irreducible components, we have $\so(7)$ and if it has three components,
the gauge algebra is $\so(8)$.  Since the prediction from field theory as well as the results we find here differ for these cases, we will treat them separately.  

We are going to argue that one needs only to consider configurations of the form $\cC_{2n_1,\dots,2n_N}\equiv[\text{I}_{2n_1}, \dots, \text{I}_{2n_N}]$.
Concerning type III and type IV, we recall from the previous section that type IV intersecting type I${}_0^\ast$ always has monodromy, {\it i.e.}, it carries
 algebra $\su(2)$, so at least as far as global symmetry is concerned, we can replace 
$[\dots, \text{IV}, \dots, \text{III},\dots] \rightarrow[\dots, \text{I}_2, \dots, \text{I}_2,\dots]$.  
Moreover, both type III and type IV yield higher values for
$\dt_P$ with respect to type I${}_2$, resulting in stronger constraints on the remainder of the configuration.
Finally, type I${}_n$ intersecting type I${}_0^\ast$ always has monodromy as summarized in table \ref{t:sumloccontr}.  Thus, type I${}_{2n}$ and type I${}_{2n+1}$ 
contribute the same algebra summand, although the former yields a smaller value for $\dt_P$.  Hence, in the remaining of this section we will restrict 
our attention to configurations of the form $\cC_{2n_1,\dots,2n_N}$ as defined above.

This has a nice additional consequence; as can be seen from the type I${}_0^\ast$ entry in table \ref{t:maxn}, the local models
are highly constrained if either $A$ or $B$ are non-zero.  Thus, we conclude that the maximal global symmetry will be realized for configurations
$\cC_{2n_1,\dots,2n_N}$ for $A=B=0$, as anticipated above.

\subsubsection*{Gauge algebra $\so(8)$}

Let us start by supposing that there is no monodromy on $\Sigma$, that is, type I${}_0^\ast$ on $\Sigma$ carries $\so(8)$ algebra. 
In this case, the monodromy cover takes the form
\begin{align}
(\psi - \alpha)(\psi-\beta)(\psi-\gamma)~,
\end{align}
where $\alpha$, $\beta$ and $\gamma$ are sections of $-2K_B-\Sigma$ and $\alpha+\beta+\gamma=0$. 
Let $\sigma$ be a transverse coordinate to $\Sigma$ in some open set, then we can represent the quantities $\alpha$, $\beta$ and $\gamma$
as polynomials of an appropriate degree in $\sigma$. The degree is determined by how many times the sections vanish (with multiplicity) on $\Sigma$.
This means that they vanish exactly 
\begin{align}
(-2K_B-\Sigma)\cdot\Sigma = 4-m
\end{align}
times on $\Sigma$.  In particular, $\deg \alpha=\deg\beta=\deg\gamma=4-m$. Now, the residual discriminant takes the form
\begin{align}
\label{eq:discrso8}
{\Delta\over z^6}\Big|_{z=0} = - (\alpha-\beta)^2(2\alpha+\beta)^2(\alpha+2\beta)^2~.
\end{align}
By construction, $\alpha-\beta, 2\alpha+\beta, \alpha+2\beta$ are also sections of $-2K_B-\Sigma$ and in particular they cannot coincide pairwise 
(otherwise the third vanishes identically or $\beta=0$). 
The maximal algebra will be achieved if we can tune the above sections such that each of the factors of \eqref{eq:discrso8} has
a (different) unique zero of multiplicity $4-m$. If that were the case, the configuration would therefore be $\cC_{8-2m,8-2m,8-2m}$,
yielding the global symmetry algebra $\sp(4-m)\oplus\sp(4-m)\oplus\sp(4-m)$,
which would match the predictions from field theory.

Unfortunately, this tuning is only possible when $4-m=1$, and cannot 
be carried out in the other cases.  
To see this, consider two homogeneous polynomials of degree $4-m$ 
on $\mathbb{CP}^1$, each with a zero of multiplicity $4-m$ (representing
two of the three factors appearing in \eqref{eq:discrso8}.  Introducing
an appropriate local coordinate $t$, we may assume that the zeros of one
polynomial are at $t=0$ and those of the other polynomial are at $t=1$.
For the third factor, then, we seek constants $c_0$ and $c_1$ such that
\begin{equation}
F(t):=c_0t^{4-m} + c_1(t-1)^{4-m} \label{eq:linearcomb}
\end{equation}
has a zero of multiplicity $4-m$ at some value of $t$ other than $0$ or $1$.
In particular, if $4-m>1$, then \eqref{eq:linearcomb} must have a multiple
root so that $F(t)$ and $F'(t)$ must share a zero (other than
$t=0$ or $t=1$).  
But since $F(t) - \frac t{4-m}F'(t) = -c_1(t-1)^{3-m}$, the only
possible location of a common zero is $t=1$, contradicting our assumption
that it could be located at a third point. This shows that $F(t)$ cannot 
have roots with multiplicity higher than one; the $4-m$ roots of $F(t)$ 
of multiplicity one could in principle support 
$4-m$ curves of type I${}_2$.

The conclusion is that we can achieve 
$\sp(4-m)\oplus\sp(4-m)\oplus\sp(1)^{\oplus(4-m)}$ symmetry.

\subsubsection*{Gauge algebra $\so(7)$}

In this case, the monodromy cover \eqref{eq:moncovI0st} takes the form
\begin{align}
\label{eq:so7moncov}
(\psi-\lambda)(\psi^2+\lambda\psi+\mu)~,
\end{align}
where $\lambda$ is a section of $-2K_B-\Sigma$ and $\mu$ is a section of $-4K_B-2\Sigma$. 
As before, we think of them as polynomials of the appropriate degree in some transverse coordinate $\sigma$ in 
some open set of $\Sigma$, and their degrees are given by
\begin{align}
\label{eq:alphabetasecs}
\deg \lambda &= (-2K_B-\Sigma)\cdot \Sigma = 4-m~,		&\deg \mu &= (-4K_B-2\Sigma)\cdot\Sigma = 8-2m~.
\end{align}
The residual discriminant takes the form
\begin{align}
\label{eq:resdiscrso7}
\Deltat\equiv {\Delta\over z^6}\Big|_{z=0} = - \left(\lambda^2-4\mu\right)\left(2\lambda^2-\mu\right)^2~,
\end{align}
where $\varphi\equiv\lambda^2-4\mu$ is not a square (otherwise the algebra is $\so(8)$). 
We notice that by construction
$\varphi$ and $2\lambda^2-\mu$ cannot have common zeros. It is possible that $\varphi$ has a factor that is a square, 
and we write $\varphi=\delta\gamma^2$, where
$\delta$ is not a square. In particular, this means that $\deg\delta$ cannot vanish. Note that for $m=3,$ necessarily $\gamma\equiv1$.
The degrees of these quantities are as follows
\begin{align}
\deg (2\lambda^2-\mu) &=8-2m~,		&8-2m=2\deg \gamma+\deg\delta &\geq 2\deg\gamma+2 ~.
\end{align}
This means that $(2\lambda^2-\mu)^2$ can support at most type I${}_{16-4m}$, while $\gamma^2$ can support at
most type I${}_{6-2m}$. The maximal algebra will be achieved if we can tune the above sections as described, 
yielding the configurations $\cC_{16-4m,6-2m}$, with algebras $\sp(8-2m)\oplus\sp(3-m)$.
These coincide with the predictions from field theory.

\subsubsection*{Gauge algebra $\g_2$}

When $\Sigma$ carries gauge algebra $\g_2,$ the Tate form prescribes the divisibility conditions
\begin{align}
\label{eq:divcondg2}
\ut &\equiv u/z~, 	&\vt &\equiv v/z^2~, 		&\wt &\equiv w/z^3~,
\end{align}
where (see appendix \ref{app:Intate})
\begin{align}
f&=-\frac13 u^2+v~,		&g&=\frac2{27}u^3-\frac13 uv +w~.
\end{align}
In terms of \eqref{eq:divcondg2} the residual discriminant reads
\begin{align}
\Deltat\Big|_{z=0} = {\Delta\over z^6}\Big|_{z=0} = 4\ut^3\wt - \ut^2\vt^2 - 18\ut\vt\wt + 4\vt^3 + 27 \wt^2~.
\end{align}
Let $\cC_{2n_1,\dots,2n_N}$ be a configuration of the type defined above, where $\Sigma_i=\{\sigma_i=0\}$ carries type I${}_{2n_i}$.
If $2\sum_i n_i = \dt_\Sigma$, then the residual discriminant must assume the form $\Deltat =c \sigma_1^{2n_1}\cdots\sigma_N^{2n_N}$, 
where $c$ is a constant. That is, $\Deltat$ is a square
and the algebra along $\Sigma$ cannot be $\g_2$. This yields a first constraint
\begin{align}
\sum_i n_i \leq \half\dt_\Sigma-1 = 11-3m~.
\end{align}
For each curve in $\cC_{2n_1,\dots,2n_N}$ (following our discussion in appendix \ref{app:Intate}), we define the quantities 
\begin{align}
U_{(i)} &\equiv \ut|_{z=0}~,		&V_{(i)} &\equiv {\vt\over \sigma^{n_i}}\Big|_{z=0}~,	&W_{(i)} &\equiv {\wt\over \sigma^{2n_i}}\Big|_{z=0}~.
\end{align}
We are going to prove the following:
\begin{lemma}
\label{l:deltattg2}
Given $\Sigma=\{z=0\}$ carrying type I${}_0^\ast$ with gauge algebra $\g_2$, any configuration $\cC_{2n_1,\dots,2n_N}$ is such that
$ n_i\leq10+3\Sigma^2$.
\end{lemma}
This follows from the global constraints on $\at_\Sigma$ and $\bt_\Sigma$
and on the degrees of $U,V,W$. Let type I${}_{2n}$ be supported along $\Sigma'=\{\sigma=0\}$ and be part of a 
configuration on type I${}_0^\ast$. First, we show that for $n > 10-3m$ the following hold
\begin{align}
\label{eq:degreesconstrg2fg}
\deg V +n &= 2\deg U~, 		&\deg W+2 n &=3\deg U~.
\end{align}
While this is trivial in the $m=1$ and $m=2$ cases, we need to be careful when $m=3$. In fact, let $m=3$ and $n=2$. Imposing
\begin{align}
\label{eq:fI0starg2}
8-2m = \deg \ft\big|_{z=0} &= \deg \left( -\frac13U^2 + V\sigma^{n}\right)~,
\end{align}
it follows that either $\deg U \leq1$ and $\deg V =0$ or $\deg V = 2\deg U - 2$. The second constraint  
\begin{align}
\label{eq:gI0starg2}
12-3m = \deg \gt\big|_{z=0} &= \deg \left( \frac2{27}U^3 -\frac13 UV\sigma^{n}+ W\sigma^{2n} \right)~,
\end{align}
instead forces 
\begin{align}
\deg W +4= 3\deg U	\qquad \text{or}	\qquad \deg W + 4 = 2 + \deg U + \deg V~.  
\end{align}
The first possibility trivially leads to $\deg U>1$. The second implies $\deg U + \deg V \geq2$, which again forces $\deg U>1$ or $\deg V>0$. 
When $n =3$ (the only other possibility for $m=3$), the claim is again trivial. Therefore, we verified that \eqref{eq:degreesconstrg2fg}
hold in all cases of interest for this lemma.

Let $m=3$. We need to show that type I${}_4$ is forbidden. This follows easily from the above since for this curve, we have $\deg U\leq1$,
but \eqref{eq:degreesconstrg2fg} forces $\deg U \geq2$. 

Let $m=2$. To show that type I${}_{10}$ is not allowed, we explicitly check that \eqref{eq:fI0starg2} and \eqref{eq:gI0starg2} cannot be both satisfied. 
Here, we have $\deg U \leq4$ but \eqref{eq:degreesconstrg2fg} (for $n=5$) fixes $\deg U =4$, $\deg V=3$ and $\deg W=2$. We expand 
these quantities accordingly as
\begin{align}
U &= u_0 + u_1\sigma + u_2\sigma^2+ u_3\sigma^3 + u_4\sigma^4~,	&V&= v_0 + v_1\sigma+ v_2\sigma^2 + v_3\sigma^3~, 		
&W&=w_0+ w_1\sigma + w_2\sigma^2~,
\end{align}
where $u_4,v_3,w_2\neq0$.
By imposing \eqref{eq:fI0starg2}, we obtain
\begin{align}
v_2 &=\frac23 u_3u_4~,		&v_3&=\frac13 u_4^2~,  	&v_1&=\frac13u_3^2 +\frac23 u_2 u_4~,
&v_0&=\frac23 u_2 u_3 - \frac23 u_1 u_4~,
\end{align}
and in order for \eqref{eq:gI0starg2} to hold, we must satisfy
\begin{align}
u_3 (u_1 u_3 - 2 u_4)&=0~,	&u_2 u_3^2 + u_4^2&=0~,		&u_3^3 + 6 u_2 u_3 u_4 + 3 u_1 u_4^2&=0~.
\end{align}
The only solution to the above system of equations is $u_3=u_4=0$, which is unacceptable.
Thus, $\deg g|_{z=0}\geq7$ contradicting \eqref{eq:gI0starg2}.

If $m=1$, we need to show that I${}_{16}$ is forbidden.
It is possible to treat this case as well as in the above, that is, by expanding $U$, $V$ and $W$ to the appropriate degrees and showing that 
$\deg f|_{z=0}=6$ and $\deg f|_{z=0}=9$ cannot both be satisfied. Since the procedure is identical but the algebra
is more involved, we do not report these calculations here. This concludes the proof of the lemma.

We can generalize the lemma above in the following way. We want to show that a configuration such that $\sum_i n_i = 11-3m$ cannot 
be supported on type I${}_0^\ast$ with $\g_2$ gauge algebra. Our argument goes as follows. Suppose $\sum_i n_i = 11-3m$, then the
residual discriminant takes the following form
\begin{align}
\label{eq:resdiscrg2F2}
\Deltat & = (\sigma_1^{n_1}\cdots\sigma_N^{n_N})^2 F_{[2]}~,
\end{align}
where $F_{[2]}$ can be represented as a polynomial of degree 2 in the local variable $\sigma$. If \eqref{eq:resdiscrg2F2} can be rewritten 
as \eqref{eq:resdiscrso7}, this means that the underlying gauge algebra must be enhanced to at least $\so(7)$. This amounts to investigating whether
we can find appropriate sections $\alpha$ and $\beta$ as in \eqref{eq:alphabetasecs}.  As it happens, we can simply count parameters. In fact,
for a configuration $\cC_{2n_1,\dots,2n_N}$ the number of parameters that we need to completely specify the form of 
\eqref{eq:resdiscrg2F2} is $N+3$, where the first summand corresponds to the locations of the $N$ transverse curves and the additional 
3 parameters characterize the polynomial $F_{[2]}$. 
We already see that the most constraining configurations will be the ``longest", that is $n_i=1$ for each $i=1,\dots,N=11-3m$.
In particular, this means that in every case
\begin{align}
\#(\Deltat) \leq 11-3m+3=14-3m~.
\end{align}
On the other hand, \eqref{eq:alphabetasecs} implies that the numbers of parameters
that specify the local sections $\alpha$ and $\beta$ are 
\begin{align}
\#(\alpha)&=4-m+1~, 	&\#(\beta)&=8-2m+1~.
\end{align}
Therefore, $\#(\alpha)+\#(\beta)\geq \#(\Deltat)$ always holds. This gives us our final bound, 
\begin{align}
\sum_i n_i \leq 10-3m~,
\end{align}
on allowed configurations.
We emphasize the fact that we did not make any statements regarding the realizability of these configurations;
we only argued that if the configuration can be consistently constructed, the algebra must be at least $\so(7)$ (this happens for example in the case $m=3$
with $\cC_{2,2}$).

With these results at our disposal, we conclude that the maximal configurations will be $\cC_{20-6m}$, yielding the algebras $\sp(10-3m)$.
These agree with the predictions from field theory.

\subsection{Type I${}_n^\ast$}

In this section, we analyze configurations on a curve $\Sigma=\{z=0\}$ carrying type I${}_n^\ast$, $n\geq1$, and 
$(\at,\bt,\dt)_{\Sigma}=(8-2m, 12-3m, 24+(n-6)m)$, where $\Sigma^2=-m$.  
(Recall that when $n\ge4$, we have $m=4$ and for I${}_3^*$ with monodromy,
we have $m=2$ or $m=4$.)
From table \ref{t:maxnstar}, we see that configurations involving curves carrying
other than type II and type I${}_{n'}$ are forbidden, and we discard the former since it does not contribute to the global symmetry.
The degrees of vanishing of $f$ and $g$ along $\Sigma$ are 2 and 3 respectively, thus the results in appendix \ref{app:Intate} for type I${}_0^\ast$
(for $A=B=0$) naturally extend here, that is, each curve I${}_{n'}$ in the configuration has monodromy and the relevant configurations are again  
$\cC_{2n_1,\dots,2n_N}$ with the associated algebras $\sp(n_1)\oplus\cdots\oplus\sp(n_N)$. 

The following result will uniquely determine the maximal length of the configuration:
\begin{result}
\label{r:atbt23}
Let $\Sigma'=\{\sigma=0\}$ be a curve carrying any Kodaira type and $P\equiv\Sigma\cap\Sigma'$.  Then $(\at_P,\bt_P)_\Sigma=(2k,3k)$ for $k=0,1,2,\dots$.
\end{result}
In fact, from appendix \ref{app:Instartate} we recall
\begin{align}
{f\over z^2}\Big|_{z=0}&=-\frac13u_1^2~,		&{g\over z^3}\Big|_{z=0}&=\frac2{27}u_1^3~,
\end{align}
where $u_1$ is a locally defined function in a neighborhood of $\Sigma$.  Let $k$ be the degree of $u_1$ in the variable $\sigma$, than 
the result immediately follows from \eqref{eq:defordP}.

We will now discuss the cases for different values of $n$.

\subsubsection*{$\bf{n<4}$ odd}

Here the residual determinant reads
\begin{align}
\label{eq:n1delta}
{\Delta\over z^{2i+5}}\Big|_{z=0}&= 4 u_1^3 w_{2i+2}~,
\end{align}
where $n=2i-1$. From appendix \ref{app:Intate} and result \ref{r:atbt23} we have $\sigma_1^{2n_1}\cdots\sigma_N^{2n_N} | w_{2i+2}$, 
and it is then convenient to define
the local function $\phi$ such that
\begin{align}
\label{eq:defphiodd}
w_{2i+2} \equiv  \phi \sigma_1^{2n_1}\cdots\sigma_N^{2n_N}~.
\end{align}
In terms of degrees, \eqref{eq:n1delta} reads
\begin{align}
\label{eq:degreesnoddless6}
\dt_\Sigma - 2\sum_i n_i = 3\deg u_1 + \deg\phi~.
\end{align}
As noted above, $\deg u_1=\at_\Sigma/2$, and the monodromy on type I${}_n^\ast$ for $n$ odd is determined by whether 
$w_{2n+2}$ has a square root. This means that $\deg \phi$ must be even for the case without monodromy (indicated by the upperscript ``s"), while
necessarily $\deg \phi>0$ for the monodromy case (indicated by the upperscript ``ns"). In order to impose the least constraint on $\sum_i n_i$,
we will then assume $\deg\phi=0$ for I${}_n^{\ast \text{s}}$ and $\deg\phi=2$ for I${}_n^{\ast \text{ns}}$. 
For $n=1$, we have the following table:
\begin{align}
\xymatrix@R=0mm@C=5mm{
m		&\dt_\Sigma		&\deg u_1			&(\sum_i n_i)^s		&\text{max. algebra}		&(\sum_i n_i)^{ns}		&\text{max. algebra} \\
1		&19				&3				&5				&\sp(5)				&4					&\sp(4)\\
2		&14				&2				&4				&\sp(4)				&3					&\sp(3)\\
3		&9				&1				&3				&\sp(3)				&2					&\sp(2)\\
4		&4				&0				&2				&\sp(2)				&1					&\sp(1)
}
\end{align}
From table \ref{t:fieldthypreds}, we read off the algebras predicted from field theory: $\sp(6-m)\oplus\sp(4-m)$ and $\sp(5-m)\oplus\sp(4-m)$
for the two monodromy cases.  We have just determined that our configurations yield algebras that are contained in the first summands.

For $n=3,$ the analysis is even simpler, as $\dt_\Sigma - 3\deg u_1 = 12$ for any value of $m$. Therefore, maximal configurations yield $\sp(6)$
in the case without monodromy and $\sp(5)$ in the monodromy case.  The prediction from field theory is $\sp(6)$ for the former case and 
$\sp(9-m)\oplus\so(2-\half m)$ for the latter (with $m=2$ or $m=4$).

\subsubsection*{$\bf{n=2}$}

The analysis for this case differs from the previous section as the residual discriminant now reads 
\begin{align}
\label{eq:n2delta}
{\Delta\over z^{8}}{\Big|_{z=0}}&=  u_1^2 (4u_1w_5-v_3^2)~.
\end{align}
Considering again a configuration of the form $\cC_{2n_1,\dots,2n_N},$ we define
\begin{align}
\label{eq:defphieven}
(4u_1w_5-v_3^2) \equiv  \phi \sigma_1^{2n_1}\cdots\sigma_N^{2n_N}~,
\end{align}
and we have the following relation
\begin{align}
\label{eq:degreesnevenless6}
\dt_\Sigma -2\sum_i n_i = 2\deg u_1 + \deg \phi~.
\end{align}
The monodromy condition is determined by whether $(4u_1w_{7}-v_{3}^2)$ admits a square root, and we rephrase 
this in terms of a (minimal) constraint on the degree of $\phi$: $\deg \phi=0$ for I${}_2^{\ast\text{s}}$ and $\deg \phi=2$ for I${}_2^{\ast\text{ns}}$.
The results are
\begin{align}
\xymatrix@R=0mm@C=5mm{
m		&\dt_\Sigma		&\deg u_1			&(\sum_i n_i)^s		&\text{max. algebra}		&(\sum_i n_i)^{ns}		&\text{max. algebra} \\
1		&20				&3				&7				&\sp(7)				&6					&\sp(6)\\
2		&16				&2				&6				&\sp(6)				&5					&\sp(5)\\
3		&12				&1				&5				&\sp(5)				&4					&\sp(4)\\
4		&8				&0				&4				&\sp(4)				&3					&\sp(3)
}
\end{align}
Again, these are subalgebras of the first summands of the field theoretical algebras from table \ref{t:fieldthypreds}.

\subsubsection*{$\bf{n\geq4}$}

As before, there are two monodromy cases here, yielding gauge algebras $\so(2n+7)$ for I${}_n^{\ast\text{ns}}$ and $\so(2n+8)$ for I${}_n^{\ast\text{s}}$.
The corresponding predictions from table \ref{t:fieldthypreds} are $\sp(2n-1)$ and $\sp(2n)$, respectively.

Since $m=4$, we have $\deg u_1=0$ and 
\begin{align}
2\sum_i n_i = \dt_\Sigma - \deg \phi = 4n - \deg \phi~,
\end{align}
where $\phi$ is defined as in \eqref{eq:defphiodd} for the odd case or as in \eqref{eq:defphieven} in the even case.
With monodromy, we have also $\deg\phi>0$ and the maximal configuration is given by $\cC_{d_\Sigma-2}$, yielding global symmetry algebra $\sp(2n-1)$; in the no monodromy case, we assume $\deg\phi=0$ as above and $\cC_{d_\Sigma}$ is the maximal
configuration with $\sp(2n)$ as global symmetry.
These both agree with the field theory predictions.

\subsection{Type IV${}^\ast$, III${}^\ast$ and II${}^\ast$}

Here the situation is quite simple.  These Kodaira types are not allowed to intersect any of the singularity types yielding a non-vanishing gauge algebra.
Therefore, there is no global symmetry coming from F-theory.  In particular, this result matches with the field theory prediction for type II${}^\ast$, for which 
no matter content is available.

\subsection{Summary} 
\label{subsec:summary}

Finally, we can summarize the results we have derived in this section.
Tables \ref{t:summary1} and \ref{t:summary2} list the (relatively) maximal symmetry algebras of F-theory constructions
of the class of 6D SCFTs we consider in this work. We emphasize once more the nature of the results we find. 
Comparing with table \ref{t:fieldthypreds}, we conclude that in some cases the global symmetry from
field theory can be realized; in other cases, this cannot happen and we determine the (relatively) maximal global symmetry algebras.

In fact, the agreement between the F-theory constructions and the Coulomb
branch predictions is quite good, and we can easily list the cases where there
is a mismatch.  Note that in some cases, there is more than one way
to produce the gauge algebra in F-theory; it is enough for our purposes
if we can find agreement between the field theory prediction and at least
one of the F-theory realizations.  With that being said, the cases
where there is a mismatch are:
\begin{enumerate}
\item In the case of $\su(2)$ with $m=2$, we have obtained
only $\so(7)$ rather than the Coulomb branch expectation of $\so(8)$.
This is the case which has already been explained in
\cite{Ohmori:2015fk}, which found a field-theoretic reason that the SCFT
should have a smaller global symmetry than the one observed on the Coulomb
branch.
\item In the case of $\su(3)$ with $m=1$, we see the most severe mismatch:
the predicted global symmetry 
is $\su(12)$, but we find only a variety of different subalgebras of this
(different ones for different realizations of the gauge algebra).
\item For $\so(8)$ with $m=1$ or $m=2$, we only realized
$\sp(4-m)\oplus\sp(4-m)\oplus\sp(1)^{\oplus(4-m)}$ rather than the predicted
$\sp(4-m)\oplus\sp(4-m)\oplus\sp(4-m)$.
\item For $\so(n)$, $9\le n\le 13$ and $m<4$, the predicted global symmetry
associated to the spinor representation is never realized.
\item For $\so(13)$ with $m=2$, only a $\sp(5)$ subalgebra of the predicted
$\sp(7)$ algebra is realized.
\item Finally, for $\mathfrak{f}_4$, $\mathfrak{e}_6$, and $\mathfrak{e}_7$,
none of the predicted global symmetries are realized.
\end{enumerate}

\begin{table}
\begin{center}
\begin{tabular}{c|c|c|c}
type	along $\Sigma$&algebra on $\Sigma$	&	$-\Sigma^2$	&max. global symmetry algebra(s) \\\hline
 \multirow{2}{*}{I${}_2$}		&\multirow{2}{*}{$\su(2)$}	&2	&$\su(4)$ \\\cline{3-4}
 &	&	\multirow{1}{*}{1}	&	$\so(20)$ \\\hline
  \multirow{4}{*}{I${}_{n\geq3}$, $n$ odd}		&\multirow{2}{*}{$\sp([n/2])$}	&\multirow{2}{*}{1}	&	$\so(13+2n)$ \\
 &	&	 													& $\so(7+2p)\oplus\so(7+2n-2p)$, $0\leq p\leq \ff{n+1}2$
 \\\cline{2-4}
 &	\multirow{2}{*}{$\su(n)$}	&2	&$\su(2n)$ \\\cline{3-4}
 &		&\multirow{1}{*}{1}	&	$\su(8+n)$ \\\hline
  \multirow{3}{*}{I${}_{n\geq4}$, $n$ even}		&\multirow{1}{*}{$\sp(n/2)$}	 &\multirow{1}{*}{1}	&	 $\so(16+2n)$ 
   \\\cline{2-4}
 &	\multirow{2}{*}{$\su(n)$}	&2	&$\su(2n)$ \\\cline{3-4}
 &		&\multirow{1}{*}{1}	&	$\su(8+n)$ \\\hline
   \multirow{1}{*}{I${}_{6}$}		&	\multirow{1}{*}{$\su(6)^\ast$}	&1	&$\su(15)$ \\\hline
\multirow{4}{*}{III}	& \multirow{4}{*}{$\su(2)$}			&2	&$\so(7)$\\\cline{3-4}
&	&	\multirow{3}{*}{1}&	$\so(7)\oplus\so(7)\oplus\su(2)$ \\
&	&				&	$\so(7)\oplus\su(6)$ \\
&	&				&	$\su(10)$ \\\hline
\multirow{9}{*}{IV} & \multirow{2}{*}{$\su(2)$} & 2	&$\so(7)$ \\\cline{3-4}
			&		&1		& 	$\so(7)\oplus\so(7)\oplus\su(3)$ \\\cline{2-4}
    & \multirow{7}{*}{$\su(3)$} 
& \multirow{1}{*}{3} & - \\ \cline{3-4}
&
& \multirow{2}{*}{2}	&$\su(3)\oplus\su(3)$\\
    & & &$\sp(2)$\\\cline{3-4}
    &	&	\multirow{4}{*}{1}&	$\su(3)^{\oplus4}$ \\
    &	&				&	$\su(3)^{\oplus2}\oplus\sp(2)$\\
&	&				&	$\su(3)\oplus\sp(3)$ \\
&	&				&	$\sp(4)$ \\\hline
 \multirow{10}{*}{I${}_0^\ast$}	& \multirow{3}{*}{$\mathfrak{g}_2$}	&	
3		&	$\sp(1)$ \\\cline{3-4}
 &	&	2		&	$\sp(4)$ \\\cline{3-4}
 &	&	1		&	$\sp(7)$ \\\cline{2-4}
 						& \multirow{3}{*}{$\so(7)$}	&	
3		&	$\sp(2)$ \\\cline{3-4}
 &	&	2		&	$\sp(4)\oplus\sp(1)$ \\\cline{3-4}
 &	&	1		&	$\sp(6)\oplus\sp(2)$ \\\cline{2-4}
						& \multirow{4}{*}{$\so(8)$}	&	4	&	-	\\\cline{3-4}
 &	&	3		&	$\sp(1)\oplus\sp(1)\oplus\sp(1)$ \\\cline{3-4}
 &	&	2		&	$\sp(2)\oplus\sp(2)\oplus\sp(1)^{\oplus2}$ \\\cline{3-4}
 &	&	1		&	$\sp(3)\oplus\sp(3)\oplus\sp(1)^{\oplus3}$ \\\hline
\end{tabular}
\end{center}
\caption{Global symmetries of F-theory models.}
\label{t:summary1}
\end{table}

\begin{table}
\begin{center}
\begin{tabular}{c|c|c|c}
type	along $\Sigma$&algebra on $\Sigma$	&	$-\Sigma^2$	&max. global symmetry algebra \\\hline
 \multirow{8}{*}{I${}_1^\ast$}	&	\multirow{4}{*}{$\so(9)$}	& 	4	&	$\sp(1)$ \\\cline{3-4}
 &	&	3	&	$\sp(2)$ \\\cline{3-4}
  &	&	2	&	$\sp(3)$ \\\cline{3-4}
   &	&	1	&	$\sp(4)$ \\\cline{2-4}
   &	\multirow{4}{*}{$\so(10)$}	& 	4	&	$\sp(2)$ \\\cline{3-4}
 &	&	3	&	$\sp(3)$ \\\cline{3-4}
  &	&	2	&	$\sp(4)$ \\\cline{3-4}
   &	&	1	&	$\sp(5)$ \\\hline
    \multirow{8}{*}{I${}_2^\ast$}	&	\multirow{4}{*}{$\so(11)$}	& 	4	&	$\sp(3)$ \\\cline{3-4}
 &	&	3	&	$\sp(4)$ \\\cline{3-4}
  &	&	2	&	$\sp(5)$ \\\cline{3-4}
   &	&	1	&	$\sp(6)$ \\\cline{2-4}
   &	\multirow{4}{*}{$\so(12)$}	& 	4	&	$\sp(4)$ \\\cline{3-4}
 &	&	3	&	$\sp(5)$ \\\cline{3-4}
  &	&	2	&	$\sp(6)$ \\\cline{3-4}
   &	&	1	&	$\sp(7)$ \\\hline
    \multirow{2}{*}{I${}_3^\ast$}	&	\multirow{1}{*}{$\so(13)$}	& 	2,4	&	$\sp(5)$ \\\cline{2-4}
   &	\multirow{1}{*}{$\so(14)$}	& 	4	&	$\sp(6)$  \\\hline
    \multirow{2}{*}{I${}_{n\geq4}^\ast$}	&	\multirow{1}{*}{$\so(2n+7)$}	& 	4	&	$\sp(2n-1)$ \\\cline{2-4}
   &	\multirow{1}{*}{$\so(2n+8)$}	& 	4	&	$\sp(2n)$  \\\hline
   \multirow{2}{*}{IV${}^\ast$}	&	\multirow{1}{*}{$\mathfrak{f}_4$}	& 	1--5	&	- \\\cline{2-4}
 						  &	\multirow{1}{*}{$\mathfrak{e}_6$}	& 	1--6	&	- \\\hline
   \multirow{1}{*}{III${}^\ast$}	&	\multirow{1}{*}{$\mathfrak{e}_7$}	& 	1--8	&	- \\\hline
   \multirow{1}{*}{II${}^\ast$}	&	\multirow{1}{*}{$\mathfrak{e}_8$}	& 	12	&	- \\\hline
\end{tabular}
\end{center}
\caption{Global symmetries of F-theory models.}
\label{t:summary2}
\end{table}

One possible resolution to this mismatch is that, as in the case of 
$\su(2)$ with $m=2$, there is some as-yet-to-be-discovered field
theoretic reason for the global symmetry algebra of the SCFT to be
smaller than the one visible on the Coulomb branch.  However, it is
also conceivable that the ``missing'' global symmetries are present
in the field theory, but cannot be realized in F-theory.

Finally, note that in this paper, we only addressed the case of gauge theories,
completely ignoring the possibilities of Kodaira fibers of types
I${}_0$, I${}_1$ and II, but those possibilities do in fact occur.
The known 6D SCFTs 
 with no gauge symmetry include the
``rank $n$ $E$-string theories'' and the (2,0) theories of type A, D,
or E.  Both of these familes of examples have a construction in M-theory which
shows the presence of an $\SU(2)$ global symmetry\footnote{This
global symmetry combines with the $\SU(2)$ $R$-symmetry of a (1,0)
SCFT to form an $\SO(4)$ symmetry which is visible from the M-theory
perspective.} (see, for example, \cite{Ohmori:2014pca}).
However, this $\SU(2)$ global symmetry has no known realization in
F-theory, except for the case of the rank $2$ $E$-string theory 
\cite{Ohmori:2014kda} and
the (2,0) $A_1$ theory (which can be obtained from the proposal in
\cite{Ohmori:2014kda} by moving to the point in the Coulomb moduli
space of the rank $2$ $E$-string theory which has an $A_1$ point).
In general, the $\SU(2)$ global symmetry has no visible action on
the matter representation on the Coulomb branch, and so will be difficult
to analyze.

\section{Conclusions and outlook}
\label{sec:final} 

In this work we have carried out a study of the global symmetries of 6D SCFTs;
in particular, we compared the global symmetries arising from field theory
considerations on the Coulomb branch of the SCFT to the 
constraints on global symmetries arising from F-theory constructions.
We focused our analysis on gauge theories 
in which the Coulomb branch has dimension one.  We found agreement
in the vast majority of cases, and observed that one of the cases where
there is disagreement has a recent field theory explanation in the literature
\cite{Ohmori:2015fk}.  These results should provide the first step in
an analysis of global symmetries in the general case.
When paired with the classification scheme of \cite{atomic}, the above results might allow one to compute the relatively maximal global symmetry algebra(s) of a given 6D SCFT from the conjecturally complete 6D SCFT dictionary found in \cite{atomic}.  Associating to a SCFT its relatively maximal global symmetry algebras could shed light on topics such as the structure of those discrete $\GU(2)$ subgroups associated to 6D SCFTs, as described in \cite{6D-SCFT}.  

To be more precise, given a 6D SCFT meeting the hypotheses of \cite{atomic}, 
it has a description in F-theory in terms of an elliptic fibration degenerating over the discriminant locus, 
which in this more general context may be a linear or `D-type' chain of $\P^1$'s 
with specified negative self-intersections obeying known restrictions.  
The action of a discrete $\GU(2)$ subgroup $\Gamma$ on the base is also allowed,
 leaving an orbifold singularity in the base, $B \cong \C^2 /\Gamma$. 
In terms of the complexity of the computation of the global symmetry algebra 
of a SCFT using an appropriate generalization of the above methods, 
one significant modification from the hypotheses of the present work is that 
one needs to consider more than a single compact curve as the discriminant locus; 
for a complete description of the rules these chains must obey, see \cite{atomic}.  
Since we can associate a given 6D SCFT of type-A \footnote{A similar procedure can be applied to the curve clusters of type-D; here we restrict our discussion to type-A configurations for simplicity.} to a linear chain of pairwise transversely intersecting compact curves $\Sigma_1, \dots ,\Sigma_m$ with specified self-intersection numbers, to compute the relatively maximal global symmetry algebra(s) of that SCFT, one might proceed in a fashion similar to that above, namely, by inspecting the set of allowed configurations of non-compact curves intersecting the configuration $\Sigma_1, \dots ,\Sigma_m$ transversely.  As above, this generalization should provide a family of relatively maximal global symmetry algebras realizable in F-theory for a 6D SCFT, each arising as the direct sum of the gauge algebras of some allowed configuration of transverse non-compact curves.

We hope to return to this question in future work.

\appendix

\section{The Tate algorithm for the I${}_n$ case}
\label{app:Intate}

In this appendix, we determine the data necessary to describe a transverse intersection 
between a Kodaira type I${}_n$ supported on a divisor 
$\{\sigma=0\}$ on the base and any other Kodaira type, which we assume to be supported on a divisor $\{z=0\}$.  
For the methods of this appendix, type I${}_{m\geq1}^\ast$ can be treated on the same footing as type I${}_0^\ast$, thus
we will just refer to the latter in the following.
For these cases, however, the analysis we present here do not suffice 
to completely solve these models, as they involve additional subtleties which we will deal with separately in the main body of the paper. 

The authors of \cite{matter1} went through a careful analysis of Weierstrass models for $\su(n)$. Here we slightly generalize their approach in order to describe models for $\sp(n/2)$ as well.  A similar approach was also employed in \cite{newTate}.  The strategy is to expand the quantities in \eqref{eq:weierstr} in powers of $\sigma$
\begin{align}
\label{eq:expansions}
f &= \sum_i f_i \sigma^i~,    &g&=\sum_i g_i\sigma^i~, &\Delta&=\sum_i \Delta_i \sigma^i~,
\end{align}
where the quantities $f_i,\ g_i$ and $\Delta_i$ can be thought of as polynomials in $z$, at least locally in a neighborhood of $\{\sigma=0\}$.
The requirement for the divisor $\{\sigma=0\}$ to carry type I${}_n$ implies that $\Delta_i=0$ $\forall i < n$.  As in \cite{matter1}, we assume that the divisor $\{\sigma=0\}$ is non singular.  This guarantees that any ring of local functions on a sufficiently small open subset of $\{\sigma=0\}$ is a unique factorization domain.

In order to obtain a type I${}_1$ singularity we need to impose
\begin{align}
\Delta_0 = 4 f_0^3 +27 g_0^2 = 0~,
\end{align}
that is, there exists a local function $\phi$ such that \footnote{Given two local functions $\phi_1$ and $\phi_2$, the notation $\phi_1 \sim \phi_2$ indicates that the two functions are identical up to powers of $\sigma$.}
\begin{align}
f_0&\sim -\ff1{48} \phi^2~, &g_0&\sim \ff1{864} \phi^3~.
\end{align}
We set $f_0= -\ff1{48} \phi^2$ and $g_0= \ff1{864} \phi^3$ and we obtain: \footnote{The choice of the numerical coefficients aims to match the conventions in the existing literature.}

\underline{\bf Summary for I${}_1$}
\begin{align}
f &= -\ff1{48}\phi^2 + f_1\sigma + O(\sigma^2)~,\nonumber\\
g &= \ff1{864}\phi^3 + g_1\sigma +  O(\sigma^2)~, \nonumber\\
\Delta &= \ff1{192}\phi^3\left( 12 g_1+\phi f_1  \right) \sigma 
+ O(\sigma^2)~.
\end{align}
The coefficients $f_i,\ g_i$ above have been redefined to absorb the terms proportional to $\sigma$ we might have introduced when substituting for $f_0$
and $g_0$.  Now, setting $\Delta_1=0$ implies
\begin{align}
g_1 \sim -\ff1{12} \phi f_1~.
\end{align}
Moreover, we have that for each $i\geq1$ there are the following standard terms in the expansion of the discriminant
\begin{align}
\Delta_i = 4f_0^2f_i + 27g_0g_i + \cdots~.
\end{align}
Hence, it turns out to be convenient to redefine
\begin{align}
\gt_i = g_i + \ff1{12} \phi f_i~, \qquad i \geq1~.
\end{align}

\underline{\bf Summary for I${}_2$}
\begin{align}
\label{eq:I2summary}
f &= -\ff1{48}\phi^2 + f_1\sigma + f_2\sigma^2+O(\sigma^3)~,\nonumber\\
g &= \ff1{864}\phi^3 -\ff1{12}\phi f_1\sigma + (\gt_2 - \ff1{12}\phi f_2)\sigma^2 +  O(\sigma^3)~, \nonumber\\
\Delta &= \ff{1}{16}\left( \phi^3 \gt_2 - \phi^2f_1^2 \right)\sigma^2 + O(\sigma^3)~.
\end{align}
To ensure that the coefficient of $\sigma^2$ in $\Delta$ vanishes, we can set 
\begin{align}
\label{eq:mudef}
\phi &\sim \mu \phi_0^2~, &f_1 &\sim \ff12\mu \phi_0\psi_1~,
\end{align} 
with $\mu$ square-free (for appropriate locally-defined functions).
Now, $\Delta_2=0$ is solved by
\begin{align}
\gt_2 = \ff14 \mu\psi_1^2~.
\end{align}

\underline{\bf Summary for I${}_3$}
\begin{align}
\label{eq:I3expl}
f &= -\ff1{48}\mu^2\phi_0^4 + \ff12 \mu\phi_0 \psi_1\sigma + f_2\sigma^2+  f_3\sigma^3 + O(\sigma^4)~,\nonumber\\
g &= \ff1{864}\mu^3 \phi_0^6 -\ff1{24} \mu^2\phi_0^3\psi_1\sigma + \ff1{4}\left( \mu\psi_1^2 -\ff13 \mu\phi_0^2 f_2\right)\sigma^2
 +\left(\gt_3 - \ff1{12}\mu\phi_0^2 f_3\right)\sigma^3+ O(\sigma^4)~,\nonumber\\
 \Delta &= \ff1{16} \mu^3\phi_0^3 \left(\phi_0^3\gt_3 - \psi_1^3 -\phi_0^2\psi_1f_2\right)\sigma^3+O(\sigma^4)~.
\end{align}

Now, in order to obtain type I${}_4$ along $\{\sigma=0\}$, we need to set
\begin{align}
\Delta_3 = \ff1{16} \mu^3\phi_0^3 \left(\phi_0^3\gt_3 - \psi_1^3 -\phi_0^2\psi_1f_2\right) =0~.
\end{align}
By writing the terms in the parenthesis as $\psi_1^3 + \phi_0(\dots)$, 
we have that $\phi_0\big|_{\sigma=0}$ must divide $\psi_1\big|_{\sigma=0}$, {\it i.e.},
\begin{align}
\psi_1 \sim -\ff13 \phi_0\phi_1~.
\end{align}
Now, $\Delta_3=0$ is solved by setting
\begin{align}
\gt_3 = -\ff13 \phi_1f_2 - \ff1{27} \phi_1^3~.
\end{align}
Again, for each $i\geq3$ we have terms in the expansion of $\Delta_i$ of the form
\begin{align}
\Delta_i = 24 f_0f_1f_{i-1} + 27g_0\gt_i + \cdots~,
\end{align}
so that it is convenient to redefine
\begin{align}
\gh_i &= \gt_i + \ff13 \phi_1 f_{i-1}~,	&\fh_2 &= f_2+\ff13 \phi_1^2~.
\end{align}

\underline{\bf Summary for I${}_4$}
\begin{align}
\label{eq:I4}
f &= -\ff1{48}\mu^2\phi_0^4 - \ff16 \mu\phi^2_0 \phi_1\sigma + (\fh_2-\ff13 \phi_1^2)\sigma^2+ f_3\sigma^3 + f_4\sigma^4 +  O(\sigma^5)~,\nonumber\\
g &= \ff1{864}\mu^3 \phi_0^6 + \ff1{72} \mu^2\phi_0^4\phi_1\sigma + \ff1{6}\left(\ff13 \mu\phi_0^2\phi_1^2 -\ff12  \mu\phi_0^2 \fh_2\right)\sigma^2 \nonumber\\
&\quad  +\left(-\ff13 \phi_1 \fh_2 + \ff2{27} \phi_1^3  - \ff1{12}\mu\phi_0^2 f_3\right)\sigma^3
+\left( \gh_4 - \ff13\phi_1f_3 - \ff1{12}\mu\phi_0^2f_4  \right)\sigma^4+ O(\sigma^5)~,\nonumber\\
 \Delta &= \ff1{16} \mu^2\phi_0^4 \left(-\fh_2^2 + \mu\phi_0^2\gh_4\right)\sigma^4+O(\sigma^5)~.
\end{align}

Next, in order for $\Delta_4$ to vanish along $\{\sigma=0\}$, we require that $\mu\phi_0\big|_{\sigma=0}$ divides $\fh_2\big|_{\sigma=0}$, {\it i.e.},
\begin{align}
\fh_2 \sim \ff12 \mu\phi_0\psi_2~,
\end{align}
for some locally defined function $\psi_2$.  We obtain type I${}_5$ along $\{\sigma=0\}$ by setting
\begin{align}
\gh_4 = \ff14\mu\psi_2^2~.
\end{align}

\underline{\bf Summary for I${}_5$}
\begin{align}
\label{eq:I5}
f &= -\ff1{48}\mu^2\phi_0^4 - \ff16 \mu\phi^2_0 \phi_1\sigma + (\ff12 \mu\phi_0\psi_2-\ff13 \phi_1^2)\sigma^2+ f_3\sigma^3 + f_4\sigma^4 +f_5\sigma^5+  O(\sigma^6)~,\nonumber\\
g &= \ff1{864}\mu^3 \phi_0^6 + \ff1{72} \mu^2\phi_0^4\phi_1\sigma + \ff1{6}\left(\ff13 \mu\phi_0^2\phi_1^2 -\ff14  \mu^2\phi_0^3 \psi_2\right)\sigma^2 \nonumber\\
&\quad  +\left(-\ff16 \mu\phi_0\phi_1 \psi_2 + \ff2{27} \phi_1^3  - \ff1{12}\mu\phi_0^2 f_3\right)\sigma^3
+\left( \ff14 \mu\psi_2^2 - \ff13\phi_1f_3 - \ff1{12}\mu\phi_0^2f_4  \right)\sigma^4 \nonumber\\
&\quad +\left( \gh_5 -\ff13\phi_1 f_4 -\ff1{12} \mu\phi_0^2 f_5 \right)\sigma^5  + O(\sigma^6)~,\nonumber\\
 \Delta &= \ff1{16} \mu^3\phi_0^4 \left(\phi_0^2\gh_5 + \phi_1\psi_2^2 -\phi_0\psi_2f_3\right)\sigma^5+O(\sigma^6)~.
\end{align}

We can go one step further and find a complete general form for type I${}_6$ by setting $\Delta_5=0$, which reads
\begin{align}
\phi_1\psi_2^2 - \phi_0\psi_2 f_3 + \phi_0^2 \gh_5 =0~.
\end{align}
Now, we factorize the roots of $\phi_0|_{\sigma=0}$ according to which ones divide $\phi_1|_{\sigma=0}$ or $\psi_2|_{\sigma=0}$, as
\begin{align}
\phi_0 &\sim \alpha\beta~, 	&\psi_2 &\sim -\ff13\alpha \phi_2~, 	&\phi_1&\sim\beta\nu~,
\end{align}
where $\alpha, \beta, \phi_2$ and $\nu$ are locally-defined functions, and $\alpha\big|_{\sigma=0}$ is the greatest common divisor of $\phi_1\big|_{\sigma=0}$ and $\psi_2\big|_{\sigma=0}$.  This implies that $\beta\big|_{\sigma=0}$ and $\phi_2\big|_{\sigma=0}$ do not share any common factor, 
and that $\Delta_5=0$ is satisfied by
\begin{align}
f_3 &\sim -\ff13 \nu\phi_2-3\beta\lambda~, 	&\gh_5&\sim \phi_2\lambda~,
\end{align}
for some locally-defined function $\lambda$. Finally

\underline{\bf Summary for I${}_6$}
\begin{align}
f &= -\ff1{48}\mu^2\alpha^4\beta^4 - \ff16\mu\alpha^2\beta^3\nu\sigma + \ff13 (-\ff12  \mu \alpha^2\beta\phi_2 -\beta^2\nu^2)\sigma^2 \nonumber\\
&\quad + (-\ff13 \nu\phi_2-3\beta\lambda)\sigma^3 + f_4\sigma^4 +f_5\sigma^5+ f_6\sigma^6+ O(\sigma^7)~,\nonumber\\
g &= \ff1{864}\mu^3 \alpha^6\beta^6 + \ff1{72} \mu^2\alpha^4\beta^5\nu\sigma + \ff1{18}\left( \mu\alpha^2\beta^4\nu^2 +\ff14  \mu^2\alpha^4\beta^3 \phi_2\right)\sigma^2 \nonumber\\
&\quad  +\left(\ff1{12} \mu\alpha^2\beta^2\nu\phi_2 + \ff2{27} \beta^3\nu^3  + \ff1{4}\mu\alpha^2\beta^3\lambda \right)\sigma^3
\nonumber\\
&\quad +\left( \ff1{36} \mu\alpha^2\phi_2^2 + \ff19 \beta\nu^2\phi_2 + \beta^2\nu\lambda- \ff1{12}\mu\alpha^2\beta^2f_4  \right)\sigma^4\nonumber\\
&\quad +\left( \phi_2\lambda -\ff13\beta\nu f_4 -\ff1{12} \mu\alpha^2\beta^2 f_5 \right)\sigma^5  
+\left( \gh_6 - \ff13 \beta\nu f_5 - \ff1{12} \mu\alpha^2\beta^2 f_6  \right)\sigma^6+ O(\sigma^7)~,\nonumber\\
 \Delta &= \ff1{432} \mu^2\alpha^4\beta^3 \left(27\mu\alpha^2\beta^3\gh_6+9\mu\alpha^2\beta^2\phi_2f_4 +\mu\alpha^2\phi_2^3\right. \nonumber\\
& \qquad\qquad\qquad\ \left.  -243\lambda^2 \beta^3+ 54\phi_2\nu\lambda\beta^2 - 3\beta\nu^2\phi_2^2 \right)\sigma^6+O(\sigma^7)~.
\end{align}

Solving algebraically $\Delta_6=0$ is a daunting challenge and we are not able to do it in complete generality.  In order to deal with type I${}_n$ for $n\geq7$,
we implement the inductive argument of  \cite{newTate}, of which we state here the results. 

\underline{\bf Summary for I${}_{i}$, $i\geq7$}
\begin{align}
\label{eq:Ininductive}
f &= -\ff13 u^2 + v~,\nonumber\\
g&= \ff{2}{27} u^3 - \ff13uv + w~, \nonumber\\
\Delta&= 4 u^3w -u^2v^2 -18uvw +4v^3+27w^2~,
\end{align}
where $u,\ v$ and $w$ are local functions such that
\begin{align}
\sigma^{[i/2]}&|\ v~,	&\sigma^{2[i/2]}&|\ w~,	&\sigma&\nmid u~.
\end{align}
In the even case, {\it i.e.}, for $i=2n$, the expansion of $u$ in powers of $\sigma$ takes the form
\begin{align}
\label{eq:uinduct}
u = \ff14 \mu\phi_0^2 + u_1\sigma + \cdots + u_{n-1}\sigma^{n-1}~,
\end{align}
where, as before, $\mu\big|_{\{\sigma=0\}}$ is square-free, while $v$ and $w$ are generic.  In the odd case, $i=2n+1$, these have instead the
expansions 
\begin{align}
\label{eq:genforsvwnodd}
v &= \ff12 \mu\phi_0 t_n \sigma^n + v_{n+1}\sigma^{n+1} + \cdots ~,\nonumber\\
w &= \ff14 \mu t_n^2 \sigma^{2n} + w_{2n+1}\sigma^{2n+1} + \cdots ~,
\end{align}
where $t_n$ is a locally defined function.  We note that this is indeed the most general solution for I${}_m$, $m\geq10$, while it is a solution, but not
the most general one, for $m<10$.  We explicitly constructed above the most general solution for $m\leq6$, so we have three cases, {\it i.e.,} $m=7,8,9$ for which 
we may not capture the full story.  In this paper we treat these cases as described by \eqref{eq:Ininductive}.

\subsubsection*{Local models for intersections}

We want to use the above expressions for type I${}_n$ in order to construct local models describing intersections of a divisor $\{\sigma=0\}$ carrying type
I${}_n$ and a divisor $\{z=0\}$ carrying any other Kodaira type.  In order to do this,
let $(a,b,d)$ be the degrees of vanishing of $(f,g,\Delta)$ along $\{z=0\}$.
Then we have the divisibility conditions
\begin{align}
z^a&|\ f~,		&z^b&|\ g~,	&z^d&|\ \Delta~.
\end{align}
Imposing these conditions on each term of the expansions \eqref{eq:expansions} leads in turn to certain divisibility conditions on the various local functions
describing the local model for the intersection.
In particular, we must check whether the point of intersection $P=\{z=\sigma=0\}$ has degrees of vanishing
equal or higher than $(4,6,12)$, in other words, if it describes a non-minimal intersection\footnote{We refer to this condition 
as the ``(4,6,12) condition".}.  If that is that case, we discard this solution.

\subsubsection*{Monodromy}

For type I${}_n$, $n\geq3$, there is a monodromy condition we need to test in order to specify the full algebra: in fact, the gauge algebra summand is either 
$\su(n)$ or $\sp([n/2])$.
The condition for monodromy is determined by the function $\mu$ defined in \eqref{eq:mudef}: we are in the case without monodromy, {\it i.e.,}~the algebra is
$\su(n)$, when $\mu\big|_{\{\sigma=0\}}$ has no zeros, and conversely we are in the monodromy case when $\mu$ vanishes somewhere along $\{\sigma=0\}$.
As we will see in the example below, there are cases where the general form of the local model for the intersection forces $\mu$ to vanish somewhere.  That is,
we can prove that in order to have an allowed intersection the monodromy on the type I${}_n$ is fixed.

\subsubsection*{An example}

Let us consider the intersection between a type I${}_4$ along the divisor $\{\sigma=0\}$ and a type I${}_0^\ast$ along the locus $\{z=0\}$.  We take the
degrees of vanishing of $f,\ g$ and $\Delta$ along $\{z=0\}$ to be 2, 3 and 6 respectively.  This means that 
\begin{align}
\label{eq:I4I0star}
z^2 &|\ f~, \quad\mbox{but }   z^3 \nmid f~,     &z^3& |\ g~, \quad\mbox{but }   z^4 \nmid g~. 
\end{align}
Imposing these constraints on each term of the solution \eqref{eq:I4} implies that 
\begin{align}
z^2&|\ f_i~,\quad i\geq3~, 		&z^3&|\ \gh_i~, \quad i\geq4~,		&z^2&|\ (\fh_2-\ff13\phi_1^2) \quad \Longrightarrow \quad z^2|\ \fh_2~,\ z|\ \phi_1
\end{align}
and either $z| \phi_0$ or $z| \mu$.  Let us distinguish between these two cases:
\begin{enumerate}
\item \underline{$z|\ \phi_0$ and $z\nmid \mu$.} Let us rewrite our quantities by factoring out the explicit partial $z$ dependence determined above; 
for simplicity of notation, we indicate the functions with the same symbol, as
\begin{align}
f &= -\ff1{48}z^4\mu^2\phi_0^4 - \ff16z^3 \mu\phi^2_0 \phi_1\sigma + (z^2\fh_2-\ff13 z^2\phi_1^2)\sigma^2+ z^2f_3\sigma^3 + z^2f_4\sigma^4 +  z^2O(\sigma^5)~,\nonumber\\
g &= \ff1{864}z^6\mu^3 \phi_0^6 + \ff1{72} z^5\mu^2\phi_0^4\phi_1\sigma + \ff1{6}z^2 \mu\phi_0^2\left(\ff13z^2\phi_1^2 -\ff12 z^2\fh_2\right)\sigma^2 \nonumber\\
&\quad  +\left(-\ff13 z^3\phi_1 \fh_2 + \ff2{27} z^3\phi_1^3  - \ff1{12}z^4\mu\phi_0^2 f_3\right)\sigma^3
+\left( z^3\gh_4 - \ff13z^3\phi_1f_3 - \ff1{12}z^4\mu\phi_0^2f_4  \right)\sigma^4+ z^3 O(\sigma^5)~.
\end{align}
The degrees of vanishing of $f$ and $g$ at $P$ are 4 and 6, respectively.  This solution is therefore non-minimal and it is not allowed.
\item \underline{$z|\mu$ and $z\nmid \phi_0$.} In this case, we have 
\begin{align}
f &= -\ff1{48}z^2\mu^2\phi_0^4 - \ff16z^2 \mu\phi^2_0 \phi_1\sigma + (z^2\fh_2-\ff13 z^2\phi_1^2)\sigma^2+ z^2f_3\sigma^3 + z^2f_4\sigma^4 +  z^2O(\sigma^5)~,\nonumber\\
g &= \ff1{864}z^3\mu^3 \phi_0^6 + \ff1{72} z^3\mu^2\phi_0^4\phi_1\sigma + \ff1{6}z \mu\phi_0^2\left(\ff13z^2\phi_1^2 -\ff12 z^2\fh_2\right)\sigma^2 \nonumber\\
&\quad  +\left(-\ff13 z^3\phi_1 \fh_2 + \ff2{27} z^3\phi_1^3  - \ff1{12}z^3\mu\phi_0^2 f_3\right)\sigma^3
+\left( z^3\gh_4 - \ff13z^3\phi_1f_3 - \ff1{12}z^3\mu\phi_0^2f_4  \right)\sigma^4+ z^3 O(\sigma^5)~,\nonumber\\
 \Delta &= \ff1{16} z^2\mu^2\phi_0^4 \left(-z^4\fh_2^2 + z^4\mu\phi_0^2\gh_4\right)\sigma^4+z^6O(\sigma^5)~.
\end{align}
Now, the degree of vanishing of $f$ and $g$ at $P$ are 2 and 3, respectively, and therefore, this is an allowed intersection.  Moreover, the fact that 
$\mu\big|_{\{\sigma=0\}}$ vanishes along the locus $\{z=0\}$ implies that we have monodromy and the gauge algebra for the type I${}_4$ is $\sp(2)$.
\end{enumerate}

\subsubsection*{Summary}
Following the procedure described above, we can summarize our results for local models describing intersections of type I${}_n$ 
with curves of even and odd type as well as
I${}_0^\ast$ (this case can be extended to I${}_{m\geq1}^\ast$).  Table \ref{t:maxn} describes the maximum $n$ such that 
type I${}_n$ is allowed by the (4,6,12) condition to intersect the singularity types listed below.
The symbol - indicates that this possibility is not realizable, for example in all odd types we cannot have $\alpha\neq0$. 
\begin{table}[!h]
\begin{center}
\begin{tabular}{c | c c c c c c c c c }
~ 			& $A=B=0$			&$A=1$		&$A\geq2$		&$B=1$		&$B\geq2$	\\
\hline
II${}^\ast$ 	& 0					& 0			& 0				& -			& - 	 \\
III${}^\ast$ 	& 0					& -			& -				& 0			& 0	 \\
IV${}^\ast$ 	& 1					& 1			& 1				& -			& -	 \\
I${}_0^\ast$ 	& $\infty$				& 3			& 3				& 2			& 2	 \\
IV		 	& $\infty$				& 3			& 3				& -			& -	 \\
III		 	& $\infty$				& -			& -				& $\infty$		& 3	 \\
II			& $\infty$				& $\infty$		& 4				& -			& -
\end{tabular}
\caption{Allowed intersections for type I${}_n$.}
\label{t:maxn}
\end{center}
\end{table}

The results summarized in this table have some immediate important consequences. First, configurations $\cC_{n_1,\cdots,n_N}$ of type I${}_{n_i}$ curves do not 
yield any non-abelian symmetry for type II${}^\ast$,  III${}^\ast$ and  IV${}^\ast$. Second, in the most interesting case of type I${}_0^\ast$, table \ref{t:maxn} provides 
strong constraints if either $A\neq0$ or $B\neq0$. In both cases, by using the monodromy result in table \ref{t:sumloccontr}, the only admissible
algebra summand is $\sp(1)$. This is trivially a sub-case of the more general $A=B=0$ setting, to which we restricted our attention 
in the main body of the paper.

For each allowed model, we can compute the minimal local contributions and in some cases, as we have seen in the example above, we can gather information 
about the monodromy.  We will restrict to the case $A=B=0$, since from the above this is the relevant case.  We restrict our attention to intersections for type I${}_n$,
$n\geq2$, since these are the ones carrying a non-abelian gauge algebra.  In table \ref{t:sumloccontr}, we used b.p.~to indicate that both monodromies on type 
I${}_n$ are possible.
\begin{table}[!h]
\begin{center}
\begin{tabular}{c | c c c c c c c c c }
~ 			& $n$				&gauge algebra&	$\at$		&$\bt$		&$\dt$	\\
\hline
I${}_0^\ast$ 	& $\geq2$				& $\sp([n/2])$ 		&0 		&0 			&6 	 \\
IV		 	& 2					& $\su(2)$			&0 		&2 			&4 	 \\
IV		 	& 3					& b.p.			&1 		&2 			&4 	 \\
IV		 	& 3					& $\su(2)$			&0 		&3 			&6 	 \\
IV		 	& 4					& $\sp(2)$			&0 		&4 			&8 	 \\
IV		 	& 5					& $\sp(2)$			&0 		&5 			&10 	 \\
IV		 	& 6					& $\sp(3)$			&0 		&6 			&12 	 \\
IV		 	& $\geq7$				& $\sp([n/2])$		&0 		&$n$ 		&$2n$   \\
III		 	& $2$				& $\su(2)$			&1		&1			&3	 \\
III		 	& $3$				& b.p.			&2		&2			&6	 \\
III		 	& $4$				& b.p.			&2		&1			&6	 \\
III		 	& $5$				& $\sp(2)$			&3		&3			&9	 \\
III		 	& $6$				& $\sp(3)$			&3		&3			&9	 \\
III		 	& $\geq7$				& $\sp([n/2])$		&$\lceil n/2 \rceil$		&$\lceil n/2 \rceil$		&$3\lceil n/2 \rceil$	
\end{tabular}
\caption{Local contributions for type I${}_n$.}
\label{t:sumloccontr}
\end{center}
\end{table}

\section{The Tate algorithm for the I${}_n^\ast$ case}
\label{app:Instartate}

In this appendix, we collect the general forms for I${}_n^\ast$ for $n\geq1$,
taken from \cite{newTate}. Let $\{\sigma=0\}$ carry type I${}_n^\ast$. We write 
the quantities $f$, $g$ and $\Delta$ as expansions in the variable $\sigma$, where the
coefficients are local functions in a sufficiently small open subset of $\{\sigma=0\}$.

\underline{\bf Summary for I${}_1^\ast$}
\begin{align}
f&=-\ff13u_1^2\sigma^2 + f_3\sigma^3+ O(\sigma^4)~,\nonumber\\
g&=\ff2{27}u_1^3\sigma^3 + (\gt_4-\ff13 u_1 f_3)\sigma^4+ O(\sigma^5)~,\nonumber\\
\Delta &= 4u_1^3\gt_4 \sigma^7 + O(\sigma^8)~, 
\end{align}
where we set $\gt_j=g_j+\ff13 u_1f_{j-1}$ for $j\geq4$.

\underline{\bf Summary for I${}_2^\ast$}
\begin{align}
f&=-\ff13u_1^2\sigma^2 + f_3\sigma^3+ f_4 \sigma^4+O(\sigma^5)~,\nonumber\\
g&=\ff2{27}u_1^3\sigma^3 -\ff13 u_1 f_3 \sigma^4+ (\gt_5-\ff13 u_1f_4)\sigma^5+O(\sigma^6)~,\nonumber\\
\Delta &= u_1^2\left( 4u_1\gt_5-f_3^2\right) \sigma^8 + O(\sigma^9)~.
\end{align}

\underline{\bf Summary for I${}_3^\ast$}
\begin{align}
f&=-\ff1{48}\mu_1^2s_0^4\sigma^2 + \ff12\mu_1s_0t_2\sigma^3+ f_4\sigma^4 + f_5 \sigma^5+O(\sigma^6)~,\nonumber\\
g&=\ff1{864}\mu_1^3s_0^6\sigma^3 -\ff1{24}\mu_1^2s_0^3t_2 \sigma^4+ \ff14\mu_1( t_2^2 -\ff13 s_0^2 f_4)\sigma^5
+ (\gt_6-\ff1{12}\mu_1 s_0^2 f_5)\sigma^6+ O(\sigma^7)~, \nonumber\\
\Delta &= \ff1{16}\mu_1^3s_0^3\left( \gt_6s_0^3-f_4s_0^2 t_2 - t_2^3\right) \sigma^9 + O(\sigma^{10})~, 
\end{align}
where $\mu_1$ is square-free.

The case $n\geq4$ is fully captured by an induction argument analogous to the one presented in the previous appendix for type I${}_n$. 

\underline{\bf Summary for I${}_{n\geq4}^\ast$}

For the even case, $n=2m$,
we have
\begin{align}
\label{eq:largeneven}
u&=u_1\sigma + u_2\sigma^2 + \cdots + u_{m}\sigma^{m}~,\nonumber\\
v&=v_{m+2}\sigma^{m+2} + v_{m+3}\sigma^{m+3} + \cdots~, \nonumber\\
w&=w_{2m+3}\sigma^{2m+3} + w_{2m+4}\sigma^{2m+4} + \cdots~,
\end{align}
while for the odd case,
$n=2m+1$, we have
\begin{align}
\label{eq:largenodd}
u&=u_1\sigma + u_2\sigma^2 + \cdots + u_{m+1}\sigma^{m+1}~,\nonumber\\
v&=v_{m+3}\sigma^{m+3} + v_{m+4}\sigma^{m+4} + \cdots~, \nonumber\\
w&=w_{2m+4}\sigma^{2m+4} + w_{2m+5}\sigma^{2m+5} + \cdots~,
\end{align}
where $f$, $g$ and $\Delta$ are as in \eqref{eq:Ininductive}.

\subsubsection*{Summary}
We can summarize our results for local models on type I${}_n^\ast$. 
Table \ref{t:maxnstar} lists the upper bounds on $n$ such that the intersection between type 
I${}_n^\ast$ with the various Kodaira singularities is not non-minimal.
Again, by the symbol - we indicate that this possibility is not realizable, while the symbol X represents a non-minimal intersection $\forall n\geq1$.
\begin{table}[!h]
\begin{center}
\begin{tabular}{c | c c c c c c c c c }
~ 				& $A=B=0$			&$A=1$		&$A\geq2$		&$B=1$		&$B\geq2$	\\
\hline
II${}^\ast$ 		& X					& X			& X				& -			& - 	 \\
III${}^\ast$ 		& X					& -			& -				& X			& X	 \\
IV${}^\ast$ 		& X					& X			& X				& -			& -	 \\
I${}_0^\ast$ 		& X					& X			& X				& X			& X	 \\
I${}_{p\geq1}^\ast$ 	& X					& -			& -				& -			& -	 \\
IV			 	& X					& X			& X				& -			& -	 \\
III			 	& X					& -			& -				& X			& X	 \\
II				& 1					& 1			& 1				& -			& -
\end{tabular}
\caption{Allowed intersection for type I${}_n^\ast$.}
\label{t:maxnstar}
\end{center}
\end{table}

\end{document}